\documentclass[%
 reprint,
 superscriptaddress,
 amsmath,amssymb,
aps,
pre,
]{revtex4-2}

\usepackage[utf8]{inputenc} 
\usepackage[T1]{fontenc}
\usepackage{amsmath}
\usepackage{amsthm} 
\usepackage{booktabs} 
\usepackage{amssymb} 
\usepackage{physics}
\usepackage{graphicx}
\usepackage[mathlines]{lineno}
\usepackage{lipsum}
\usepackage{dcolumn}
\usepackage{bm}

\usepackage[pdfstartview=FitH,
            breaklinks=true,
            bookmarksopen=false,
            bookmarksnumbered=true,
            colorlinks=true,
            linkcolor=black,
            citecolor=black,
            urlcolor=black,
            pdftitle={TCRinfo},
            pdfauthor={James Henderson et al.},
            ]{hyperref}

\begin{document}

\title{Limits on Inferring T-cell Specificity from Partial Information}

\author{James Henderson}
 \affiliation{Division of Infection and Immunity}
 \affiliation{Institute for the Physics of Living Systems}
\author{Yuta Nagano}
 \affiliation{Division of Infection and Immunity}
\affiliation{Division of Medicine}
\author{Martina Milighetti}
 \affiliation{Division of Infection and Immunity}
\affiliation{Cancer Institute \\
University College London}
\author{Andreas Tiffeau-Mayer}
\affiliation{Division of Infection and Immunity}
\affiliation{Institute for the Physics of Living Systems}

\begin{abstract}
A key challenge in molecular biology is to decipher the mapping of protein sequence to function. To perform this mapping requires the identification of sequence features most informative about function.
Here, we quantify the amount of information (in bits) that T-cell receptor (TCR) sequence features provide about antigen specificity. We identify informative features by their degree of conservation among antigen-specific receptors relative to null expectations. We find that TCR specificity synergistically depends on the hypervariable regions of both receptor chains, with a degree of synergy that strongly depends on the ligand. Using a coincidence-based approach to measuring information enables us to directly bound the accuracy with which TCR specificity can be predicted from partial matches to reference sequences. We anticipate that our statistical framework will be of use for developing machine learning models for TCR specificity prediction and for optimizing TCRs for cell therapies. The proposed coincidence-based information measures might find further applications in bounding the performance of pairwise classifiers in other fields.
\end{abstract}

\keywords{TCR, immune repertoire, information theory, Renyi information}

\maketitle

Mapping the amino acid sequence of a particular T-cell receptor (TCR) to its antigen specificity is a holy grail of systems immunology \cite{dash2017quantifiable,Glanville2017IdentifyingSpecificity,hudson2023can}. The T-cell receptor endows T-cells with the ability recognize snippets of pathogenic material presented on the surface of antigen presenting cells by major histocompatibility complexes (MHC) \cite{davis1988t}. TCRs are specific, meaning a given T-cell will only activate in response to a select range of antigen stimuli. Coverage of the vast antigen space explored by evolving pathogens is enabled by immense sequence variation within the TCR  \cite{rossjohn2015t,mayer2015well}, in particular within six hypervariable loops of the heterodimeric receptor, named complementary determining regions (CDRs). 

The immense diversity of TCRs implies that many have no experimentally determined ligands \cite{goncharov2022vdjdb}. Emerging computational approaches predict the specificity of such orphan TCRs by their sequence similarity to annotated TCRs \cite{dash2017quantifiable,Pogorelyy2019FrameworkAnnotation,Chronister2021TCRMatchPredicting,hudson2023can}. However, which level of partial matching is sufficient for reliable prediction has remained unclear. Moreover, there is substantial interest in understanding for which immunological questions knowledge of paired receptor chains obtainable by single-cell sequencing is worth the trade-off with the higher throughput achievable by bulk sequencing \cite{VALKIERS2022100009}, and which TCR features are most informative for machine learning applications \cite{hudson2023can, ghoreyshi2023quantitative}.

\begin{figure*}
\includegraphics[width=0.65\textwidth]{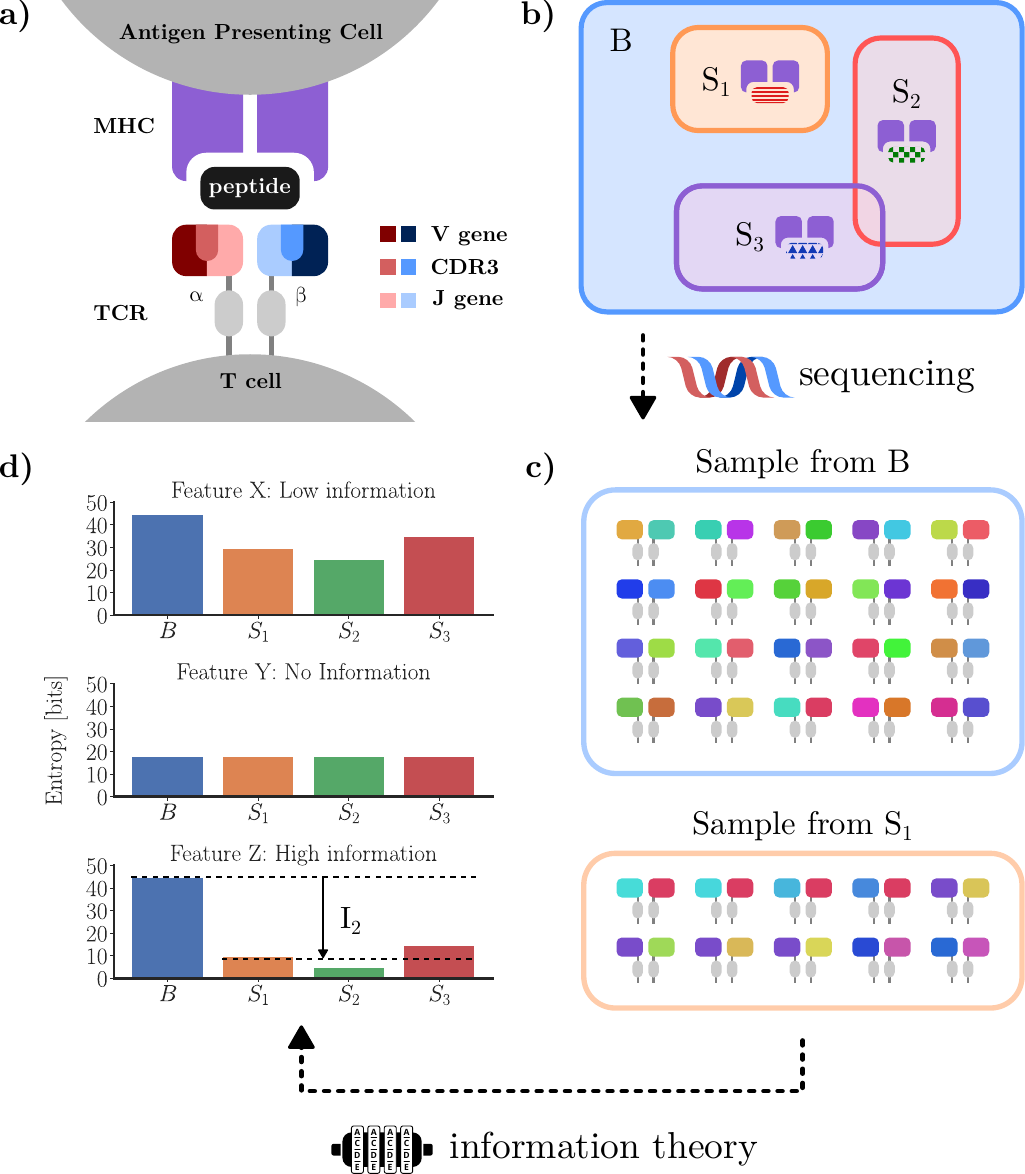}
\caption{ \textbf{Overview of analysis methodology.}  \textbf{a)} Sketch of T-cell receptor structure highlighting the V, CDR3 and J regions and their interaction with MHC-bound peptides. The TCR is composed of two chains, most commonly $\alpha$ and $\beta$ chains. Each chain in turn is comprised of a V (variable), J (joining) and C (constant) gene, with the addition of a D (diversity) gene in the $\beta$ chain. Within each chain, the CDR1 and CDR2 amino acid loops are coded for by the V gene while the CDR3 regions are at the V(D)J intersection, which is additionally diversified through the random insertion and deletion of nucleotides at gene template junctions. \textbf{b)} An abstracted view of TCR sequence space. The set B includes all possible TCRs. The subsets S$_i$ represent TCRs specific to particular ligands. \textbf{c)} Sequencing TCR from either the whole repertoire or epitope-specific subsets gives us samples from their respective distributions. \textbf{d)} The number of pairs which match in a particular feature may then be recorded to compute a probability of coincidence. The logarithm of the probability of coincidence gives a measure of the entropy of the feature. Our information theoretic approach quantifies the change in entropy between background TCRs and sets of specific TCRs of different features (top to bottom). Features which experience a large reduction in entropy (bottom) are the most informative for predicting the epitope specificity of a sequence.}
\label{tcrstruct}
\end{figure*}

Here, we address these important open questions by putting universal limits on the accuracy with which TCR specificity can be predicted from partial information. Our work takes inspiration from a long history of successful applications of information theory to the study of complex biological input-output relationships from neural coding \cite{laughlin1981simple,brenner2000synergy,palmer2015predictive} and transcriptional regulation \cite{kinney2010using,belliveau2018systematic} to pattern formation during embryo development \cite{dubuis2013positional,petkova2019optimal,mcgough2024finding}. Following recent applications of information theory to TCR repertoires by us \cite{milighetti2023analysis} and others \cite{xu2023entropic}, our analysis builds on a fundamental insight from evolutionary biology: patterns of sequence conservation in protein families provide clues about functionally relevant properties. In the immunological context, this means that TCR features that are important for specific recognition of a particular epitope will be most highly conserved among epitope-specific TCRs relative to their global diversity (Fig.~\ref{tcrstruct}).

In our current work, we provide the first comprehensive map of how much information each section of the paired chain TCR sequence provides about its specificity. To provide such a map, we make use of two recent datasets that have sequenced TCRs specific to a dozen viral MHC class I epitopes \cite{dash2017quantifiable,minervina2022sars}. We overcame statistical limitations of prior analyses of pairs of residues \cite{milighetti2023analysis,xu2023entropic, boughter2023integrated} using coincidence-based measures of repertoire diversity \cite{mayercallan}. These measures can be estimated from smaller samples than traditional measures based on Shannon entropy \cite{ma1981calculation,nemenman2011coincidences,tiffeau2023unbiased}. 
The information-theoretic approach naturally allowed us to identify synergies between different TCR sections in determining antigen specificity. Importantly, our quantification of coincidence information is underpinned by theory that directly links achievable classification accuracy to the coincidence information gained from a partial match and prior beliefs about the prevalence of epitope-specific T cells in a repertoire.

\section{An information-theoretic approach to T cell specificity}

\subsection{Coincidence analysis for features}

We have recently introduced a coincidence-based statistical framework to measure antigen-driven selection in TCR repertoires \cite{mayercallan}. The main idea of this work was to quantify clonal convergence by counting how often pairs of independently recombined clonal lineages in a sample have TCRs that are more similar to each other than some threshold level. Here, we pursue a conceptually related but novel approach that considers near-coincidences as coincidences on the level of coarse-grained TCR features. A feature may be a gene segment choice at a given locus, an amino acid at a particular residue, or a physical property of a hypervariable loop such as its charge or length. Features may also contain other features such as the $\alpha$ chain containing the V$\alpha$, J$\alpha$, and the CDR3$\alpha$ as component features.

Mathematically, a feature is a random variable that maps the sample space of all TCR sequences to a discrete set of possible categories. We denote the distribution of feature values for randomly drawn TCRs from a repertoire by $P(X)$. The probability that two independent draws return the same outcome, i.e. the probability of coincidence of $X$, is then defined by
\begin{equation}
    p_C[P(X)] = \sum_{x}P(x)^2,
\end{equation}
where $P(x)$ represents $P(X=x)$ and the sum runs over all possible outcomes of $X$.
We recall that in ecology $p_C[P(X)]$ is referred to as the Simpson's diversity index of $X$ with $D(X) = 1/p_C[P(X)]$ being an effective number of distinct species in a population \cite{simpson1949}. Intuitively, we expect the most informative features to be those whose diversity is most reduced among TCRs specific to the same epitope when compared with background TCRs. In the following we will make this intuition mathematically precise using a coincidence-based formulation of information theory.

\subsection{Coincidence entropy}

A central quantity in information theory is entropy. The entropy of a probability distribution $P(X)$ is given in its form proposed by Shannon in 1948 as \cite{shannon1948mathematical}
\begin{equation}
    \label{shannon}
    H[P(X)] = \sum_{x}P(x)\log P(x).
\end{equation}
Entropy represents the average amount of information lacking about the outcome of a measurement of random variable $X$. It is usually calculated with the logarithm taken to base $2$ such that its units are in bits and all logarithms in this following should be understood as logarithms taken with respect to this base. In 1961, Renyi showed that by relaxing one of the Shannon-Khinchin axioms from which the mathematical form of entropy is uniquely derived (strong additivity), a more general expression for entropy may be obtained \cite{renyi1961measures, khinchin1957mathematical}
\begin{equation}
    \label{renyi}
    H_{\alpha}[P(X)] = \frac{1}{1-\alpha}\log \left( \sum_{x} P(x)^{\alpha}\right),
\end{equation}
where $\alpha$ is referred to as the order of the Renyi entropy. The family of Renyi entropies include Shannon's entropy measure as the limit of $\alpha \to 1$.

We may note that the probability of coincidence introduced in the previous subsection provides a measure for the Renyi entropy of order $\alpha = 2$
\begin{equation}
\label{renyi2}
   H_{2}[P(X)] = -\log p_C[P(X)] .
\end{equation}
The Renyi entropy of order 2 is known as collision entropy in cryptography and may also be motivated from an optimal code length perspective with non-linearly weighted length penalties \cite{campbell1965coding}. Here we use the term {\it coincidence entropy} to stress its relation to coincidence-counting among sample pairs.

 \subsection{Coincidence mutual information}

We have previously used the coincidence ratio $p_C[P(X|\Pi)]/p_C[P(X)]$ between specific and and background TCRs as a measure of antigen-driven selection \cite{mayercallan}, where $p_C[P(X|\Pi)]$ is the probability of coincidence among epitope-specific TCRs averaged over a collection of epitopes, $\Pi$, and $p_C[P(X)]$ the probability of coincidence among background TCRs. Different definitions of conditional Renyi entropy for $\alpha\neq 1$ have been proposed. Here we follow \cite{jizba2004world, ilic2014generalized} and define
\begin{equation}
    \label{renyicond2-v2}
    H_{2}[P(X|Y)] = -\log p_C[P(X|Y)],
\end{equation}
where $p_C[P(X|Y)]$ is an average of $p_C[P(X|y)]$ over all outcomes of $Y$
\begin{equation}
    \label{pccond}
    p_C[P(X|Y)] = \sum_{y} \rho_{2} (y) p_C[P(X|y)],
\end{equation}
with weighting factors
\begin{equation}
    \label{weights}
    \rho_{2}(y) = \frac{P(y)^2}{\sum_{y}P(y)^2}.
\end{equation}
Detailed justification for these definitions is provided in Appendix \ref{app1}. This definition allows us to express the coincidence probability ratio in terms of coincidence entropies
\begin{equation}
    \log \left(\frac{p_C[P(X|\Pi)]}{p_C[P(X)] }\right)  = H_{2}[P(X)] - H_{2}[P(X|\Pi)].
\end{equation}
We note that for Shannon entropy this difference defines the mutual information between $X$ and $\Pi$ \cite{shannon1948mathematical}, which motivates the following definition of \emph{coincidence mutual information}
\begin{equation}
\label{mieq}
    I_{2}(X, \Pi) = \log \left( \frac{p_C[P(X|\Pi)]}{p_C[P(X)]} \right).
\end{equation}

Importantly, our definition of conditional entropy maintains additivity $   H_{2}[P(X,Y)] = H_{2}[P(X)] + H_{2}[P(Y|X)]$, where $P(X,Y)$ is the joint distribution of the random variables $X$ and $Y$. As a correlate it follows that coincidence mutual information is symmetric, $I_2(X,Y) = I_2(Y, X)$ -- as is its Shannon counterpart -- so it tells us not only how much information we gain about sequence features upon learning their epitope specificity, but also, by symmetry, how much information a sequence feature provides about its epitope specificity. Coincidence mutual information thus provides a natural way to score the importance of a TCR feature in determining specificity, which we will refer to as the feature \emph{relevancy}.

\subsection{Describing the interactions between features with redundancy and synergy}

The connection between coincidence analysis and information theory naturally allows us to apply additional notions from information theory \cite{Vergara2014, williams2010nonnegative} to describe how multiple features work in tandem to provide antigen specificity. First, \emph{conditional mutual information}
\begin{equation}
        \label{conmiu}
        I_2(X, \Pi|Y) = H_2[P(X|Y)] - H_2[P(X|\Pi,Y)],
\end{equation}
describes the remaining information provided by feature $X$ given that the value of a second feature $Y$ is already known. Here, $H_2[P(X|\Pi,Y)]$ indicates conditioning on both epitope specificity and feature $Y$. If $I_2(X, \Pi|Y) = 0$ then we refer to $X$ as a fully \emph{redundant} feature in the context of $Y$. As a trivial example, knowledge of the complete primary sequence of the full $\alpha$ chain makes any information provided by CDR3$\alpha$ redundant, and so on.

Second, \emph{interaction information}
\begin{equation}
\label{syn}
   I_{2, \mathrm{int}}(X,Y | \Pi) = I_2\left([X, Y],\Pi\right) - I_2(X,\Pi)
   -I_2(Y,\Pi)
\end{equation}
describes how much additional information both features provide in conjunction (Appendix \ref{appsynergy}). Here, $I_2\left([X, Y],\Pi\right)$ is the relevancy of the feature produced by combining the two features $X$ and $Y$. If $I_{2, \mathrm{int}}(X,Y | \Pi) > 0$ then there is \emph{synergy} between the two features. 

\section{Bounding classification accuracy of partial TCR matches}

\subsection{Pairwise classification odds}

There are well-known connections between information measures and achievable classification errors both in the Shannon \cite{Cover2005ElementsInformation} and Renyi case \cite{ben1978renyi,csiszar1995generalized}. In the following we derive how TCR classification accuracy using partial feature matches with a reference sequence is bounded when only partial information is available. We consider a classification setting, where the task is to identify spiked-in TCR sequences specific to a particular epitope $\pi$ in an otherwise naive repertoire. We will derive how posterior classification odds depend on feature relevancy and prior beliefs, i.e. the fraction of spiked-in sequences $P(\pi)$. Mathematically, in this setting the presence of a TCR sequence $\sigma$ is due to either of two generative processes:
\begin{equation}
    P(\sigma) = P(\pi) P(\sigma | \pi) + P(B) P(\sigma|B),
\end{equation}
where $P(B)=(1-P(\pi))$, and where $P(\sigma | \pi)$ describes the distribution of TCR sequences specific to epitope $\pi$ and $P(\sigma|B)$ the distribution of background TCR sequences according to V(D)J recombination. \par

To recapitulate the empirical procedure of matching TCR sequences to a database of known binders, we consider the following one-shot classification strategy: We classify a query sequence as having been generated from $P(\sigma | \pi)$, if it matches in a feature $X$ with a reference sequence randomly drawn from $P(\sigma | \pi)$. Using the odds formulation of Bayes' theorem, we may express the posterior odds of correct classification as
\begin{equation}
    \frac{P(\pi|x=x^\prime)}{P(B|x=x^\prime)} = \frac{P(x=x^\prime|\pi)}{P(x=x^\prime|B)}\frac{P(\pi)}{P(B)}.
\end{equation}
Here, $P(x=x^\prime|\pi) = p_C[P(X|\pi)]$ is the probability of a match in feature $X$ if both sequences were truly drawn from distribution $P(\sigma | \pi)$, while $P(x=x^\prime|B)$ is the probability of a match in feature $X$ for a query drawn from $P(\sigma | B)$ and a reference drawn from $P(\sigma | \pi)$. Under the assumption that the propensity of a TCR for specific binding is independent of its recombination probability \cite{mayercallan}, one can show that $P(x=x^\prime|B) = p_C[P(X)]$ (Appendix \ref{app2}). Therefore
\begin{equation}
\label{eqodds_single}
    \frac{P(\pi|x=x^\prime)}{P(B|x=x^\prime)} = \frac{p_C[P(X|\pi)]}{p_C[P(X)]}\frac{P(\pi)}{P(B)}
\end{equation}
This expression can be generalized for mixtures of multiple epitope groups, in which case the average odds over epitopes (Appendix \ref{app3}) can be expressed as
\begin{equation}
    \left<\frac{P(\pi|x=x^\prime)}{P(B|x=x^\prime)}\right>= \frac{p_C[P(X|\Pi)]}{p_C[P(X)]}\left<\frac{P(\pi)}{P(B)}\right>,
\end{equation}
where $p_C[P(X|\Pi)]$ is the conditional probability of coincidence defined previously and the averages for the odds are taken over $P(\pi)/(1-P(B))$. By the definition of coincidence mutual information (Eq. \ref{mieq}) we can rewrite the last equation as
\begin{equation}
\label{eqodds}
    \mathbb{O}_{\text{post}} = 2^{I_2(X,\Pi)} \, \mathbb{O}_{\text{prior}},
\end{equation}
which links the average posterior odds $\mathbb{O}_{\text{post}}$ to average prior odds $\mathbb{O}_{\text{prior}}$ via coincidence mutual information. Each bit of coincidence mutual information between $X$ and $\Pi$ corresponds to a two-fold gain in posterior odds. 

\begin{figure*}
\includegraphics[width=0.65\textwidth]{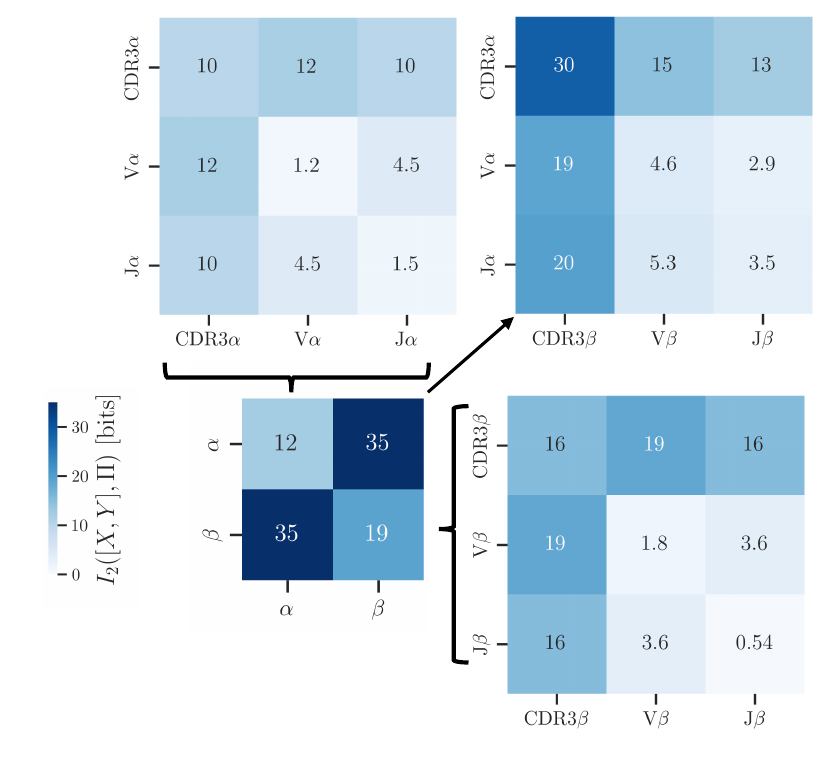}
\caption{\textbf{Coincidence mutual information between TCR sections and antigen specificity.} Relevancy scores of various sections of the T-cell receptor sequence. The off-diagonal values indicate the amount of coincidence information that combinations of features provide. The top right hand grid shows the relevancy of combination of features where one is from the $\alpha$ chain and the other the $\beta$ chain. Interaction information and conditional mutual information between features can be computed by taking the difference between the off-diagonals and the sum of the corresponding diagonal values. In particular, positive interaction information is observed between the $\alpha$ and $\beta$ chains and the CDR3 and V regions indicating synergy between these features while negative interaction information is seen between the CDR3 and J regions indicating redundancy. }
\label{results_1}
\end{figure*}

\subsection{When is partial information sufficient?}

Eq.~\ref{eqodds} captures an important Bayesian intuition about classification: Correct classification depends not only on how much information we have available, but also on our prior belief. Here, our prior belief about the likelihood that any particular sequence is specific should reflect the total fraction of spiked-in sequences. If we are searching for a needle in a haystack, this is when $\mathbb{O}_{\mathrm{prior}}$ is small, we need to use more highly informative features for correct classification. Mathematically, a minimal prior odds of $2^{-I_2(X,\Pi)}\, T$ is needed to ensure that the average posterior odds exceeds a threshold value $T$. 

Expressed in terms of prior probabilities
\begin{equation}
\label{fifty-fifty}
     P_{\text{prior}}(I_2) \geq \frac{T\, 2^{-I_2}}{1 + T\, 2^{-I_2}},
\end{equation}
is needed if only $I_2$ bits are available for classification.
To illustrate this result, we performed in-silico simulations with a toy model of TCR specificity (Appendix \ref{simulation}). These simulations showed close agreement between predicted values for $P_{\mathrm{prior}}(I_2)$ and those obtained through numerical simulation (Fig.~\ref{classification_results}).

Note that sequences drawn from $P(\sigma | B)$ may also be specific to $\pi$. Therefore, $P(\pi)$ and $P(\pi|x=x^\prime)$ are not exactly equal to the fraction of sequences specific to $\pi$ and the posterior probability of specificity, respectively. However, as shown in Appendix \ref{app4} in most cases of practical interest, where $P(\pi)$ exceeds the background frequency of sequences specific to a given epitope, this distinction is irrelevant.

\section{Application of the methodology to TCR sequence data}

We applied our framework to a curated set of multimer-sorted TCRs from CD8$^+$ T cells with specificity to viral antigens (described in detail in Appendix~\ref{methods}). We combined TCRs specific to nine SARS-CoV-2 epitopes from \citet{minervina2022sars} with TCRs specific to three epitopes from other viruses epitopes from \citet{dash2017quantifiable} (Table \ref{raw_data}). To obtain a background TCR dataset we randomly paired TCR$\alpha$ and TCR$\beta$ sequences generated by a computational model of VDJ recombination \cite{sethna2019olga}.

\subsection{A decomposition of TCR specificity into its component parts}

To provide a top-down decomposition of the information content of the TCR, we computed the relevancy of different sections of the TCR for its specificity, as well as their combinations (Fig.~\ref{results_1}). We first analyzed the information provided by the $\alpha$ and $\beta$ chains alone, which recapitulated the expected greater relevancy of the $\beta$ chain (19 bits) than the $\alpha$ chain (12 bits). By Eq.~\ref{fifty-fifty} the information provided by each chain bounds prior probabilities needed for accurate classification using single chain matches. A $\beta$ chain match requires a prior probability $P_{\mathrm{prior}} \geq 3 \cdot 10^{-5}$ for a 95\% posterior confidence. In contrast, an $\alpha$ chain match allows reliable classification only for prior probabilities $P_{\mathrm{prior}} \geq 3 \cdot 10^{-3}$. We then broke down the two chains further into their component V and J gene segments and CDR3 amino acid sequence. A CDR3$\beta$ match provides 16 bits of information (corresponding to $P_{\mathrm{prior}} \geq 4 \cdot 10^{-4}$) while a CDR3$\alpha$ match provides only 10 bits of information (corresponding to $P_{\mathrm{prior}} \geq 1 \cdot 10^{-2}$). \par 

To assess variations in feature relevancy across epitopes, we defined \emph{local relevancy} as the information gain for a specific epitope $\pi$,
\begin{equation}
\label{local_mieq}
    i_{2}(X, \pi) = \log \left( \frac{p_C[P(X|\pi)]}{p_C[P(X)]} \right).
\end{equation}
Local relevancy scores revealed a broadly consistent hierarchy of feature relevancy across epitopes (Figs.~\ref{results_2} -
\ref{results_jb}). The analysis also identified variability in local relevancy of features between different epitopes not explained by finite sampling deviations alone in line with our prior findings on a subset of the studied epitopes \cite{tiffeau2023unbiased}. We will analyze this variability in more detail in section \ref{secvariability}.

\begin{figure}
\includegraphics[width=0.8\columnwidth]{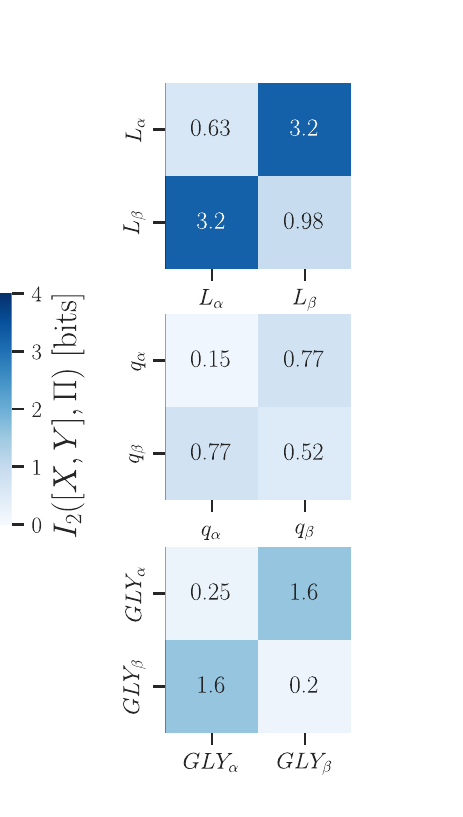}
\caption{\textbf{Coincidence mutual information between physical properties of the TCR sequence and antigen specificity.} Relevancy scores of CDR3 length, CDR3 net charge and glycine content computed for the $\alpha$ and $\beta$ chains taken independently and combined. Although each feature has modest relevancy when considered independently, these features all display substantial synergy demonstrating how physical complementarity underlies overall chain pairing constraints. }
\label{results_length_charge_gly}
\end{figure}

\subsection{CDR3 length, net charge and glycine content as features}

In addition to analysing features such as the CDR3, V and J regions, we wished to show that our definition of a feature may extend to physical properties of the TCR. To illustrate this, we computed the coincidence mutual information for the length of the CDR3 loop, its net charge and its glycine content (Fig.~\ref{results_length_charge_gly} and Figs.~\ref{results_3} - \ref{results_gly}), all features that had been described in the literature as being important for epitope specifcity \cite{dash2017quantifiable,yu2019comparative}. Our results confirm that each of these summary physical properties of the CDR3 has some relevancy in determining TCR specificity. For instance, CDR3$\beta$ net charge is roughly as informative as J$\beta$ choice. However, no individual CDR3 property captures a substantial proportion of CDR3 information demonstrating the contribution of higher order sequence features to specific binding.

\begin{figure}
\includegraphics[width=0.9\columnwidth]{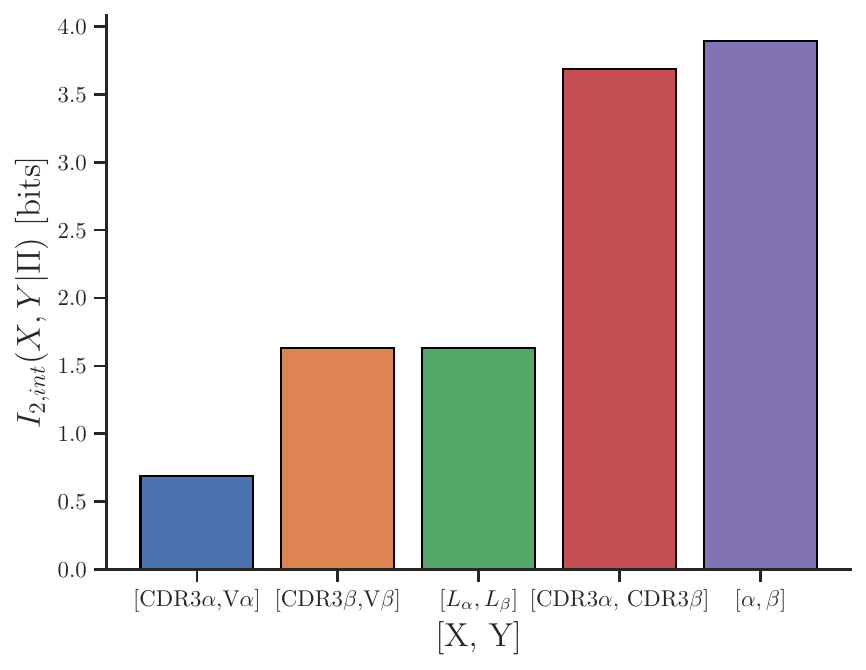}
\caption{\textbf{Synergistic TCR sequence features.} Interaction information scores for combinations of features computed from Figures \ref{results_1} and \ref{results_length_charge_gly}. Positive interaction information indicates that two features become more informative in the context of one another and hence have synergy.}
\label{synergy}
\end{figure}

\subsection{Synergy and redundancy between TCR features}

Comparing relevancy scores for individual and combined features revealed the pervasiveness of interactions between TCR sections (Fig.~\ref{results_1}) and CDR3 features (Fig.~\ref{results_length_charge_gly}), where their combined information differed from the sum of their individual relevancies. Fig.~\ref{synergy} summarizes the interaction information between important TCR features.

Our analysis identified substantial synergy between the $\alpha$ and $\beta$ chains (4 bits). This synergy implies that there are pairing restriction between $\alpha$ and $\beta$ chains in specific TCRs, which make each chain more informative when considered in its full paired chain sequence context (Appendix \ref{appsynergy}). These results broaden our prior findings to a broader set of epitopes \cite{mayercallan}, and add to a growing literature investigating TCR $\alpha$-$\beta$ pairing rules \cite{yu2019comparative, shcherbinin2020comprehensive, dupic2019genesis,carter2019single,mayercallan,milighetti2023analysis}. Pairing restrictions imply that the diversity of TCRs responding to a given epitope is lower than the product of the diversities of responsive $\alpha$ and $\beta$ chains.

We also analysed the interaction information between the CDR3 of each chain and the corresponding V segment choice. We again identified substantial synergy, presumably reflecting spatial constraints between V-gene encoded framework and CDR1/2 variability and CDR3 choice. In contrast, the interaction information between the CDR3 and J gene is negative (Figure \ref{results_1}). This is expected as the sequence variability provided by the J gene is contained in the CDR3 region \cite{lefranc2003imgt}, but demonstrates how our framework can identify redundant features without such a prior knowledge. We furthermore discovered substantial synergy between the lengths of CDR3$\alpha$ and CDR3$\beta$ regions. Similarly, glycine-content -- which is thought to be involved in CDR3 loop flexibility \cite{yu2019comparative} -- also showed synergy across chains. Collectively, these findings identify some of the key physical constraints that underlie the observed global $\alpha$-$\beta$ pairing restrictions.

\begin{figure*}
\includegraphics[width=\textwidth]{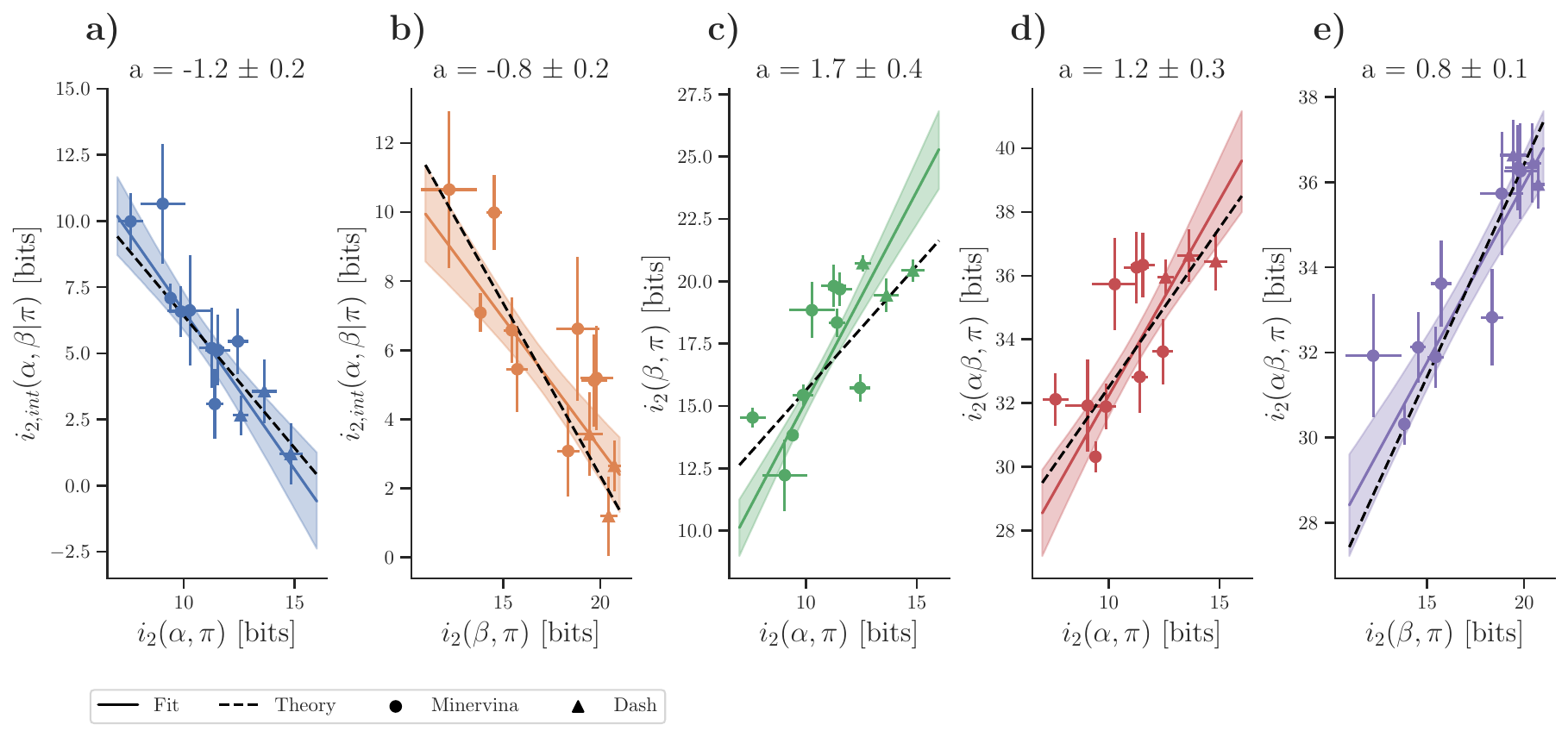}
\caption{\textbf{Correlation between $\alpha$-$\beta$ interaction information and per-chain information across epitopes.} Local interaction information and single chain information across epitopes. Weighted linear fits (solid lines) obtained using orthogonal distance regression were used to quantify the dependence between variables, with regression slopes $a$ displayed above each panel.  Epitope-specific interaction information depends negatively on the local informational value of the \textbf{a)} $\alpha$ chain and \textbf{b)} $\beta$. We furthermore find that the \textbf{c)} per-chain relevancies are positively correlated with each other as is \textbf{d,e)} total information with both single chain relevancies. The observed dependencies between variables agree well with theoretical expectations from a mixture model (dashed lines), in which epitopes differ in the number of distinct binding solutions or contain false positives.}
\label{results_5}
\end{figure*}

\subsection{Variability in interaction information across epitopes is explained by mixture models}
\label{secvariability}

To better understand potential sources of variability of TCR sequence restriction across epitopes we defined additional measures of local sequence variation: \emph{Local conditional mutual information} $i_2(X, \pi|Y) = H_2[P(X|Y)] - H_2[P(X|\pi,Y)]$ and \emph{local interaction information} $i_{2,\text{int}}(X,Y | \pi) = i_2\left([X, Y],\pi\right) - i_2(X,\pi)-i_2(Y,\pi)$. We then analysed dependencies across four variables (Fig.~\ref{results_5}): Interaction information $i_{2,\text{int}}(\alpha, \beta | \pi)$, $\alpha$-chain relevancy $i_2(\alpha,\pi)$, $\beta$-chain relevancy $i_2(\beta, \pi)$ and paired chain relevancy $i_2\left([\alpha, \beta],\pi\right)$. These analyses highlighted strong dependencies between the variables. The more informative an $\alpha$ chain or $\beta$ chain is for a given epitope, the less $\alpha$-$\beta$ interaction information contributes to global diversity restriction (Fig.~\ref{results_5}a,b). Moreover, epitopes with more informative $\alpha$ chains also have more informative $\beta$ chains (Fig.~\ref{results_5}c) and more informative full TCR sequences (Fig.~\ref{results_5}d).

Unexpectedly, all variables were highly correlated with each other and well-fitted by linear regressions, suggesting the existence of a single underlying degree of freedom that drives the observed variability across epitopes. Based on the clustering of epitope-specific TCRs we had previously proposed mixture of motif models \cite{mayercallan}, in which epitope-specific TCRs are composed of a number of distinct binding solutions (binding modes or motifs). We asked whether variability in the number of such motifs across epitopes might provide the common degree of freedom explaining the observed correlations. Deriving the expected theoretical relationships between variables (Appendix \ref{app5}), we found an increased local interaction information for epitopes with more binding modes and a decrease in individual feature relevance. Across all variable pairs studied in Fig.~\ref{results_5} the mixture model predicted linear relations with slopes of $\pm 1$, in good agreement with the best fit lines to the empirical data. Intuitively, if an epitope has multiple binding solutions, more $\alpha$ and $\beta$ chains will be able to bind it, given the right complementary chain (thus lowering the information from each individual chain). At the same time, where many solutions exist a high degree of $\alpha$-$\beta$ pairing is expected as most $\alpha$ chains from one binding solution would not be valid with $\beta$-chains from another solution (thus increasing the observed synergy between the two chains). 

Given the low prevalence of epitope-specific TCRs in a repertoire, we expect the dataset to be a mixture containing some false positive TCRs with no or low affinity to the epitope of interest even if sorting is highly specific. As we show in Appendix \ref{app6} variations in the proportion of false positives across epitopes provide an alternative explanation of the observed dependencies among variables, with high interaction information for epitopes with many false positives. Both models share the common underlying insight that epitope-specific repertoires are mixtures rather than draws from a unimodal distribution -- future research might elucidate the contributions of the different underlying mechanisms to the observed variability.

\begin{figure}
\includegraphics[width=\columnwidth]{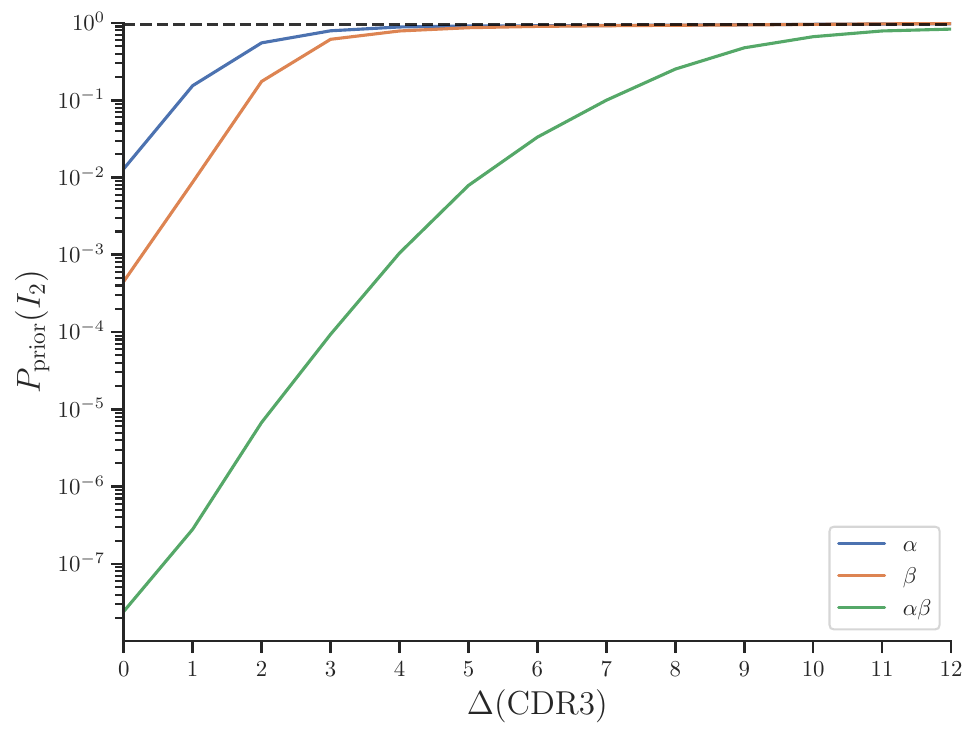}
\caption{\textbf{Information theoretic analysis of fuzzy CDR3 matches.}  Critical prior probabilities for 95\% confidence in classification using a fuzzy match with a Levenshtein distance $\Delta$. Distances are CDR3$\alpha$ and CDR3$\beta$ amino acid edit distances as well as the sum of CDR3$\alpha$+CDR3$\beta$ edit distance. Levenshtein distances are defined as the minimum number of insertions/deletions and substitutions required to turn one sequence into another.}
\label{critical_value_delta}
\end{figure}

\section{Distance metrics and near-coincidence entropy}

\subsection{Generalization of coincidence mutual information to fuzzy matches}
As exact matches are rare for complex features, it is of interest to also quantify the information provided by fuzzy feature matches.- As previously explored in Ref.~\cite{mayercallan}, we are not limited to computing the probability of exact coincidences between features but can also consider near-coincidences according to some distance metric. Given a feature $X$ distributed according to $P(X)$ and a distance metric $d(x,x^\prime)$ between outcomes $x$ and $x^\prime$, the probability that two draws from $P(X)$ are at distance $d(x,x^\prime) = \Delta$ can be defined as
\begin{equation}
    p_C[P(X)](\Delta) = \sum_{x,x^\prime} P(x)P(x^\prime) \delta_{d(x,x^\prime), \Delta},
\end{equation}
where $\delta_{d(x,x^\prime), \Delta}$ is the Kronecker delta. We use this measure to propose a \emph{near-coincidence entropy} $H_2[P(X)](\Delta) = -\log p_C[P(X)](\Delta)$, and a \emph{near-coincidence conditional entropy} $H_{2}[P(X|Y)](\Delta) = -\log p_C[P(X|Y)](\Delta)$, where $p_C[P(X|Y)](\Delta)$ once again is an average of $p_C[P(X|y)](\Delta)$ over outcomes of $Y$ using the $\rho_2(y)$ weighting factor. We define a \emph{near-coincidence mutual information}
\begin{equation}
    I_2\left(\Delta_{X,X^\prime}, Y\right) =  H_{2}[P(X)](\Delta) -  H_{2}[P(X|Y)](\Delta),
\end{equation}
where $\Delta_{X,X^\prime}$ denotes that this information is computed for near-coincidences in feature $X$ at distance $\Delta$. As this measure deals with pairs of instances of a random variable rather than single instances, this quantity cannot be defined straightforwardly for Shannon entropy but is motivated naturally when using coincidence entropies.

\subsection{Pairwise classification using fuzzy matches}

To obtain an interpretation of near-coincidence entropy, we turn once again to pairwise classification. We consider the same classification procedure as previously, but based on a fuzzy match where the sequence with unknown specificity is distance $\Delta$ from the sequence with known specificity such that $d(x,x^\prime) = \Delta$.
Similarly to our prior derivations, we find (Appendix \ref{app-fuzzy})
\begin{equation}
\label{eqodds_delta}
    \mathbb{O}_{\text{post}} = 2^{I_2\left(\Delta_{X,X^\prime},\Pi\right)} \, \mathbb{O}_{\text{prior}}.
\end{equation}
One bit of near-coincidence mutual information again corresponds to an average two-fold increase in posterior classification odds.  As in the case of exact matching, Eq. \ref{eqodds_delta} defines regimes in which fuzzy matches at a given distance are expected to succeed or fail. In Figure \ref{critical_value_delta}, we provide an example of this by computing the required prior fraction of specific sequences to obtain a posterior probability of $0.95$  with fuzzy CDR3 matches at a certain Levenshtein distance. Inversely, at a given prior odds ratio and target posterior odds ratio we can use these results to compute a critical distance beyond which classification becomes unreliable.

\section{Discussion}

The ubiquity of information theory lies in its ability to describe complex relationships between data points using a simple quantitative vocabulary. As shown by Shannon \cite{shannon1948mathematical}, entropy provides the most natural measure of uncertainty and hence changes in entropy directly capture how knowledge of one event increases understanding of another. The application of information theory to the problem of immune receptor specificity has proved highly fruitful in the past. In particular, estimates of residue Shannon entropy aided in identifying potential complementary determining regions of the TCR and immunoglobulin and highlighted that TCRs were the more diverse of these two antigen receptors \cite{stewart1997shannon}. Other, more recent studies have employed concepts from information theory such as mutual information to quantify interactions between various sections of the TCR sequence \cite{milighetti2023analysis,xu2023entropic}. These previous studies have taken a `bottom-up' approach, computing an upper bound on sequence diversity by summing up the entropy of each constituent amino acid residue or pairs of residues. In part, this `bottom-up' approach has been required due to biases in estimating Shannon entropy in small samples. Although there exist methods for reducing bias in Shannon entropy estimation, these still require resolving higher order distribution moments or essentially resort to coincidence counting and use Renyi entropy to approximate Shannon's \cite{chao2013entropy, nemenman2001entropy}. In this work, we have proposed a `top-down' approach to decomposing TCR specificity firmly rooted in second order Renyi entropy. 

Our methodology provides a general framework to assess the role of individual TCR sequence features in determining antigen specificity as well as combinations of features by introducing the concepts of relevancy, redundancy and synergy. We first compute the entropy of the full TCR sequence, divide this into its two constituent amino acid chains and then further sub-divide these into their V, J and CDR3 regions. Our results identify the $\beta$ chain as the most informative of the heterodimeric TCR's chains and the CDR3 regions to be the most informative regions of each chain. However, we also find that the information these constituent parts provide is far smaller than that of the full TCR sequence. Although these results are unsurprising, with previous work highlighting the higher contribution of the $\beta$ chain in epitope binding predictions and the importance of paired chain data \cite{springer2021contribution, boughter2023integrated, meysman2023benchmarking}, we provide the first full quantification of the information contained within these regions and, as our methodology has its foundations in coincidence-based statistics, we are able to directly interpret information measures in terms of achievable pairwise classification accuracy. Our work thus paves the way for the development of principled Bayesian methods for interpreting partial sequence matches.

Our results provide clear guides for when a limited amount of TCR sequence information, such as a single chain, is enough to solve an epitope specificity classification problem and when this loss of information may seriously impact predictive performance. We expect these insights to be important for experimental design, to decide whether the time and cost trade-off of single cell sequencing over bulk are worth the increase in information paired chain information might provide.

We have also shown how the vocabulary of information theory can be applied to TCR near-coincidence analysis, which we have introduced in recent work \cite{mayercallan}. Our framework predicts pairwise classification performance when using fuzzy matches at a given threshold TCR distance. This approach may be used to define relevant data regimes in which current or future distance metrics \cite{ehrlich2021swarmtcr, dash2017quantifiable} may be usefully applied and allows setting critical distances for classifying or clustering sequences \cite{Mayer-Blackwell}. 

Our `top-down' approach allows us to compute interaction information, which describes synergistic and redundant relationships between TCR sequence features. We observe positive interaction information, synergy, between the $\alpha$ and $\beta$ chain as well as the CDR3 and V regions, while knowledge of CDR3 regions makes their associated J regions redundant. In addition, we highlight the flexibility of our methodology by assessing the information provided by CDR3 length, net charge and glycine content. We find positive synergy between the lengths of the CDR3$\alpha$ and CDR3$\beta$ amino acid chains and their glycine content. We furthermore show how the relationship between interaction and single chain information across epitopes is compatible with a model in which epitopes vary in the number of distinct binding solutions (or possibly in the rate of false positives). With the steady accumulation of data on more epitopes, we envisage that our approach will help decipher principles underlying sequence space organization of responding TCRs.

The next steps for applying our theoretical approach are numerous. On the practical side, we propose completing the `top-down' approach and performing an analysis of the informational value of the CDR3 sequences residue by residue. This may allow for the identification of informationaly dense regions of the CDR3s and for a quantification of more complex allosteric interactions present across the receptor structure. Further extensions of our framework could account for the hierarchy of selective processes shaping the TCR repertoire by varying the background used to compute background entropy. For example, to bound the performance of multi-class classification between a set of known epitopes, it may be more appropriate to quantify the entropy of TCRs across the chosen epitope-specific groups. Likewise, sequence statistics in a naive T cell repertoire could be used as background to account for the imprint of thymic selection. Our information theoretical tools may also be used on problems other than epitope specificity. For example, one may apply them to the study of TCR-HLA associations \cite{ortega2024learning,zahid2024large} or TCR sequence to phenotype relationships \cite{carter2019single,schattgen2022integrating,ceglia2022tcri,textor2023machine}.

Linking feature information to classification isn't a problem unique to the field of protein function, nor is the task of class prediction from pairwise comparisons. Transformer neural networks, the architecture underlying the current rise of large language models, embed data in high dimensional vector spaces \cite{vaswani2017attention} and may be trained in a pairwise contrastive manner, such that items from the same class are closer together than items from different classes \cite{sun2019learning, singh2023contrastive, gao2021simcse, neelakantan2022text}. More generally, metric and representation learning commonly utilise pairwise measures to tackle problems ranging from sentence embeddings to facial recognition \cite{musgrave2020metric, bellet2013survey, wang2019multi, chen2020simple, wang2020understanding}. Our pairwise coincidence information measure may be applicable for identifying interpretable informative features in these applications.

To conclude, we have introduced a theoretical framework for mapping the information content of the T-cell receptor sequence with regards to its antigen specificity. Our results confirm prior insights from more limited structural studies regarding the relative importance of the $\alpha$ and $\beta$ chains \cite{garcia1996alphabeta,reiser2003cdr3,rossjohn2015t,Glanville2017IdentifyingSpecificity,he2020peptide, milighetti2021predicting}, but also highlight unexpected variability in the synergy between chains across epitopes. As dataset sizes continue to increase the proposed framework will be able to guide the development of machine-learning models for predicting TCR specificity from sequence by providing means to find interpretable physical features and performance bounds where information is limited. \newline

{\bf Acknowledgements.} The authors thank Benny Chain, Curtis Callan, Lisa Wagar, Naresha Saligrama, and Sankalan Bhattacharyya for useful discussions over the course of this work. The work of JH and ATM was supported in parts by funding by the Wellcome Leap HOPE Program and the Royal Free Charity. YN and MM were supported by Cancer Research UK studentships under grants BCCG1C8R and A29287, respectively.

\bibliographystyle{apsrev4-1}
\bibliography{bibliography.bib}

\clearpage
\onecolumngrid
\appendix

\setcounter{figure}{0}
\renewcommand{\thefigure}{S\arabic{figure}}
\renewcommand{\theHfigure}{S\arabic{figure}} 

\section{Defining conditional coincidence entropy}\label{app1}

We follow Refs.~\cite{jizba2004world, ilic2014generalized} and define Renyi conditional entropies as
\begin{equation}
    \label{renyicond}
    H_{\alpha}[P(X|Y)] = f^{-1}\left(\sum_{{y}} \rho_{\alpha} (y) f\left(H_{\alpha}[P(X|y)]\right)\right),
\end{equation}
where $\rho_{\alpha} (y)$ represents a generalised weighting given by
\begin{equation}
    \rho_{\alpha} (y) = \frac{P(y)^{\alpha}}{\sum_{y}P(y)^{\alpha}}.
\end{equation}
For $\alpha \neq 1$, $f$ can be any invertible function positive in $[0, \infty)$ which is a linear transform of
\begin{equation}
    f(x) = 2^{(1-\alpha)x}.
\end{equation}

Using this definition of conditional entropy ensures that entropy is additive \cite{jizba2004world,ilic2014generalized}:
\begin{equation}
    H_{\alpha}[P(X, Y)] = H_{\alpha}[P(X)] + H_{\alpha}[P(Y|X)],
\end{equation}
where $H_{\alpha}[P(X, Y)]$ represents the joint Renyi entropy of two random variables defined as
\begin{equation}
    \label{renyijoint}
    H_{\alpha}[P(X, Y)] = \frac{1}{1-\alpha} \log\left( \sum_{x, y} P(x, y)^\alpha \right).
\end{equation}
For simplicity, we will use the following definition for the conditional Renyi entropy of order $\alpha = 2$
\begin{equation}
    \label{renyicond2}
    H_{2}[P(X|Y)] = -\log \left(\sum_{y} \rho_{2} (y) 2^{-H_{2}[P(X|y)]}\right).
\end{equation}
Comparing this expression to the probability of coincidence suggests the following definition for the conditional probability of coincidence averaged over $Y$
\begin{equation}
    p_C[P(X|Y)] = \sum_{y}\rho_{2} (y) p_C[P(X|y)],
\end{equation}
such that:
\begin{equation}
    H_{2}[P(X|Y)] = -\log p_C[P(X|Y)].
\end{equation}
This definition appears quite natural as it ensures that the conditional probability of coincidence behaves like a regular conditional probability
\begin{equation}
    p_C[P(X|Y)] = \frac{p_C[P(X, Y)]}{p_C[P(Y)]},
\end{equation}
and follows Bayes' theorem
\begin{equation}
    p_C[P(X|Y)] = \frac{p_C[P(Y|X)]p_C[P(X)]}{p_C[P(Y)]}.
\end{equation}

\section{Relations between interaction information and conditional mutual information}
\label{appsynergy}
   
Interaction information and conditional mutual information can be related using the additive property of entropy to give
\begin{align}
\label{syntocondmi}
    I_{2, int}(X,Y | \Pi) &= I_2(X,\Pi|Y) - I_2(X, \Pi) \\
    &= I_2(Y,\Pi|X) - I_2(Y, \Pi).
\end{align}
Therefore, if $I_{2, int}(X,Y | \Pi) > 0$ then $X$ and $Y$ are more relevant when considered in the context of one another than when taken independently. Conversely, for $I_{2, int}(X,Y | \Pi) < 0$ features are partially redundant and becomes less informative in the context of one another. Interaction information may also be expressed in terms of the coincidence mutual information between $X$ and $Y$ to give
\begin{align}
\label{syntocondmi2}
    I_{2, int}(X,Y | \Pi) &= I_2(X,Y|\Pi) - I_2(X, Y) \\
    &= I_2(Y,X|\Pi) - I_2(Y, X).
\end{align}
Features with positive interaction information are those which become more informative about one another in the context of $\Pi$. Features with zero interaction information experience no change in their mutual information, while features with negative interaction information have a decrease in their mutual information. Finally, conditional mutual information may be expressed as
\begin{equation}
\label{syntocondmi3}
     I_2(X, \Pi|Y) = I_2(X,\Pi) + I_{2, int}(X, Y |\Pi).
\end{equation}
So that the information $X$ provides about $\Pi$ in the context of $Y$ is equal to the information it provides on its own plus a coupling term determined by its interactions with $Y$. If one feature makes another fully redundant, e.g. $I_2(X, \Pi|Y) = 0$, then the interaction information between them is $I_{2, int}(X,Y | \Pi) = -I_2(X,\Pi)$ such that their coupling completely negates the information provided by $X$ on its own. As explored in \cite{williams2010nonnegative}, definitions of redundancy and synergy may be built from interaction information and conditional mutual information. Interaction information captures a mixture of synergy and redundancy and is positive if synergy outweighs redundancy, negative if redundancy outweighs synergy and zero if they are both equal. 

\section{Linking pairwise classification odds to coincidence information}

\subsection{Likelihood ratio in terms of coincidence probabilities} \label{app2}

We begin with the posterior odds
\begin{equation}
    \frac{P(\pi|x=x^\prime)}{P(B|x=x^\prime)} = \frac{P(x=x^\prime|\pi)}{P(x=x^\prime|B)}\frac{P(\pi)}{P(B)}.
\end{equation}
Similarly to \cite{mayercallan} we may relate the probability distributions associated with $\pi$ and $B$ using the following expression: $P(x|\pi) = Q^{\pi}(x)P(x|B)$, where $Q^{\pi}(x)$ is a selection factor which reweighs the probability of each feature outcome according to some fitness function. To ensure normalisation of $P(X|\pi)$, we require that $\left<Q^{\pi}(x)\right>_{P(x|B)} = 1$. $\left<\right>_{P(x|B)}$  indicates an expectation value over the distribution $P(X|B)$. $P(x=x^\prime|\pi)$ is the probability of a match in feature $X$ if both sequences were truly drawn from distribution $P(\sigma | \pi)$. We compute this by summing over all possible ways to obtain such a match
\begin{equation}
    P(x=x^\prime|\pi) = \sum_{x, x^\prime} \delta_{x, x^\prime} P(x|\pi)P(x^\prime|\pi) = p_C[P(X|\pi)],
\end{equation}
where $\delta_{x, x^\prime}$ is the Kronecker delta.  $P(x=x^\prime|B)$ is the probability of a match in feature $X$ for a query drawn from $P(\sigma | B)$ and a reference drawn from $P(\sigma | \pi)$. This may be written as
\begin{equation}
    P(x=x^\prime|B) = \sum_{x, x^\prime} \delta_{x, x^\prime} P(x|\pi)P(x^\prime|B).
\end{equation}
Using the fact that $P(x|\pi) = Q^{\pi}(x)P(x|B)$ this may be rewritten as
\begin{equation}
    P(x=x^\prime|B) = \sum_{x, x^\prime} \delta_{x, x^\prime} Q^{\pi}(x)P(x|B)P(x^\prime|B),
\end{equation}
which may also be written as
\begin{align}
    P(x=x^\prime|B) &= \left< \sum_{x^\prime} \delta_{x, x^\prime} Q^{\pi}(x)P(x^\prime|B) \right>_{P(x|B)} \notag \\
     &= \left< Q^{\pi}(x) P(x|B) \right>_{P(x|B)}.
\end{align}
Assuming that the background probability of a particular feature is independent of its selection factor, we may decompose the expectation value to arrive at
\begin{align}
    P(x=x^\prime|B) &= \left< Q^{\pi}(x)  \right>_{P(x|B)} \left< P(x|B)  \right>_{P(x|B)} \notag \\
    &= p_C[P(X|B)]
\end{align}
As $\left< Q^{\pi}(x)  \right>_{P(x|B)} = 1$ by normalisation. We will write $p_C[P(X|B)] = p_C[P(X)]$ as it is the probability of coincidence over a background distribution.

\subsection{Average classification odds over subsets of specific TCRs}\label{app3}
The odds ratio for a particular epitope subset given a match in feature $X$ is:
\begin{equation}
    \frac{P(\pi|x=x^\prime)}{P(B|x=x^\prime)} = \frac{p_C[P(X|\pi)]}{p_C[P(X)]}\frac{P(\pi)}{P(B)}
\end{equation}
Taking the average of this over the normalised distribution of epitope subsets $P(\Pi)/(1- P(B))$ yields
\begin{equation}
\sum_{\pi}
    \frac{P(\pi|x=x^\prime)}{P(B|x=x^\prime)} \frac{P(\pi)}{1- P(B)} = \sum_{\pi} \frac{p_C[P(X|\pi)]}{p_C[P(X)]}\frac{P(\pi)}{P(B)} \frac{P(\pi)}{1- P(B)}.
\end{equation}
Recalling the definition of the conditional probability of coincidence, Eq. \ref{pccond},
the right hand side becomes
\begin{equation}
    \sum_{\pi} \frac{p_C[P(X|\pi)]}{p_C[P(X)]}\frac{P(\pi)}{P(B)} \frac{P(\pi)}{1- P(B)} = \frac{p_C[P(X|\Pi)]}{p_C[P(X)]}\frac{\sum_{\pi^\prime} P(\pi^\prime)^2}{P(B)(1- P(B))},
\end{equation}
we therefore obtain in full:
\begin{equation}
\sum_{\pi}
    \frac{P(\pi|x=x^\prime)}{P(B|x=x^\prime)} \frac{P(\pi)}{1- P(B)} = \frac{p_C[P(X|\Pi)]}{p_C[P(X)]} \frac{\sum_{\pi^\prime} P(\pi^\prime)^2}{P(B)(1- P(B))},
\end{equation}
which may be written as:
\begin{equation}
    \left<\frac{P(\pi|x=x^\prime)}{P(B|x=x^\prime)}\right> = \frac{p_C[P(X|\Pi)]}{p_C[P(X)]}\left<\frac{P(\pi)}{P(B)}\right>.
\end{equation}
Where $\left< \right>$ denotes an average over $\frac{P(\pi)}{1- P(B)}$.

\subsection{Relating the probability of specificity and the probability of being drawn from a specific subset}\label{app4}

We show here how $P(\pi)$ and $P(\pi|x=x^\prime)$, which represent the prior and posterior probabilities that a sequence is observed because it was drawn from distribution $P(\sigma|\pi)$ may be related to the prior and posterior probabilities that a sequence is in-fact specific to epitope $\pi$. We consider the mixture distribution of sequences
\begin{equation}
    P(\sigma) = P(\pi) P(\sigma | \pi) + P(B) P(\sigma|B).
\end{equation}
We denote the fraction of sequences in this mixture which would bind to epitope $\pi$ in a test of specificity as $P(S^{\pi})$. We may express this fraction in terms of the mixture proportions
\begin{equation}
    P(S^{\pi}) = P(\pi)P(S^{\pi}|\pi) + P(B)P(S^{\pi}|B),
\end{equation}
where $P(S^{\pi}|\pi)$ and $P(S^{\pi}|B)$ are the fractions of sequences produced by distributions $P(\sigma|\pi)$ and $P(\sigma|B)$ which are specific to $\pi$ respectively. As $P(\sigma|\pi)$ represents the distribution of sequences specific to $\pi$, by definition $P(S^{\pi}|\pi) = 1$, so
\begin{equation}
    P(S^{\pi}) = P(\pi) + P(B)P(S^{\pi}|B).
\end{equation}
As $P(\pi) + P(B) = 1$, this may be written as
\begin{equation}
    P(S^{\pi}) = P(\pi) + (1-P(\pi))P(S^{\pi}|B).
\end{equation}
If we assume that the fraction of sequences generated by distribution $P(\sigma|B)$ which are specific to $\pi$ is much smaller than the proportion $P(\pi)$ in our mixture so $P(S^{\pi}|B) \ll P(\pi)$, then we may simplify this directly down to:
\begin{equation}
    P(S^{\pi}) \approx P(\pi)
\end{equation}
Now substituting $P(S^{\pi})$ into our prior odds in Eq. ~\ref{eqodds_single}, we obtain
\begin{equation}
    \frac{P(\pi|x=x^\prime)}{P(B|x=x^\prime)} = \frac{p_C[P(X|\pi)]}{p_C[P(X)]} \frac{P(S^{\pi})}{1-P(S^{\pi})}.
\end{equation}
Following similar reasoning, we may write the posterior probability that a sequence sampled from the mixture is specific to epitope $\pi$ given that it matches in feature $X$ with sequence with known specificity to $\pi$ as
\begin{equation}
    P(S^{\pi}|x=x^\prime) = P(\pi|x=x^\prime)P(S^{\pi}|\pi, x=x^\prime) + P(B|x=x^\prime)P(S^{\pi}|B, x=x^\prime).
\end{equation}
Once again, $P(S^{\pi}|\pi, x=x^\prime) = 1$ and $P(S^{\pi}|B, x=x^\prime) \leq 1$, therefore
\begin{equation}
P(S^{\pi}|x=x^\prime) \approx P(\pi|x=x^\prime)
\end{equation}
as long as the posterior odds exceeds one. Taken together, this allows us to re-express Eq.~\ref{eqodds_single} in terms of prior and posterior odds of specificity to $\pi$
\begin{equation}
    \frac{P(S^{\pi}|x=x^\prime)}{1-P(S^{\pi}|x=x^\prime)} = \frac{p_C[P(X|\pi)]}{p_C[P(X)]} \frac{P(S^{\pi})}{1-P(S^{\pi})}.
\end{equation}

\subsection{Simulating pairwise classification with limited information.}\label{simulation}

\subsubsection{Generating background and specific TCRs}

Here, we followed a previously described procedure for the in silico simulation of epitope-specific repertoires \cite{mayercallan}. In short, we simulated a synthetic background repertoire by randomly generating strings  of length $k$ from a predefined alphabet of $q$ characters. We defined `epitope specific' TCRs by a set of hard coded rules: For each epitope we generated $M$ motifs, where each motif contained $c$ out of $q$ letters for each of the TCR's $k$ `residues'. To generate the specific TCRs for a motif, we produced all possible combinations of these amino acids. The set of TCRs specific to a given epitope was the full set of TCRs produced by all motifs associated with the epitope. 

\subsubsection{Comparing theoretical perforance to true classifcation accuracy}

\begin{figure}[h!]
\includegraphics[width=0.6\textwidth]{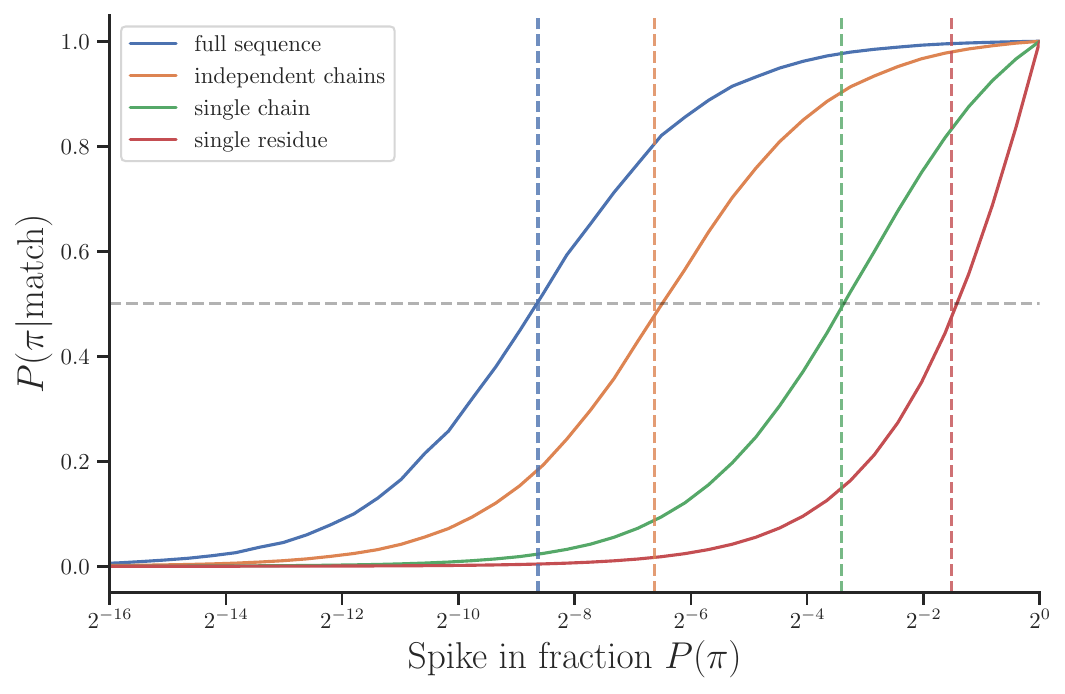}
\caption{\textbf{Simulated prior vs posterior probabilities given feature match} Prior probability of a sequence being present in a mixture distribution due to being sampled from a spiked in epitope specific distribution, vs posterior probability given a feature match with a specific sequence. Dashed lines indicate theoretical prior probability values for which a posterior probability of $0.5$ should be observed given the informational value of a given feature computed from Eq. \ref{fifty-fifty}. Solid curves cover classification for a given fraction of spiked in specific sequence in the following scenarios: using matches in the full simulated TCR (single chain), using matches in one half of the TCR (single chain), using matches from a single amino acid (single residue), using matches in both halves of the TCR but considered independently (independent chains).}
\label{classification_results}
\end{figure}

We computed the relevancy of various `features' of simulated sequences for $q=20$, $k=4$, $M=5$ and $c=3$. To define features we divided each TCR by its residues. For example, to produce an $\alpha$ and a $\beta$ chain we divided each TCR into two halves. We then produced a mixture data set into which we mixed a particular fraction of simulated background and specific TCRs. For each TCR we also assigned a label identifying which of these two original sets it came from. We then simulated pairwise classification using feature matching by sampling sequences from the mixture set and looking for matches in a particular feature with a TCR from the true set of specific sequences. Using the labels, we were able to calculate the probability that the sampled sequence was truly specific given that this match occurred. We performed this simulation using a range of features and mixture ratios as shown in Figure \ref{classification_results} and simulation results aligned with the theoretical predictions from coincidence information.

\subsection{Classification odds given a fuzzy match} \label{app-fuzzy}
For both query and reference being drawn from $P(\sigma | \pi)$, the probability of a match is
\begin{equation}
    P(d(x,x^\prime)=\Delta|\pi) = \sum_{x,x^\prime} \delta_{d(x, x^\prime), \Delta} P(x|\pi)P(x^\prime|\pi) = p_C[P(X|\pi)](\Delta).
\end{equation}
Now consider the case where the query is truly from $P(\sigma | B)$ while the reference is from $ P(\sigma | \pi)$
\begin{equation}
     P(d(x,x^\prime)=\Delta|B) = \sum_{x,x^\prime} \delta_{d(x, x^\prime), \Delta} P(x|B)P(x^\prime|\pi).
\end{equation}
Similarly to the case for exact matching, we may write $P(x^\prime|\pi) = Q^{\pi}(x^\prime)P(x^\prime|B) $ such that
\begin{equation}
    P(d(x,x^\prime)=\Delta|B) = \sum_{x,x^\prime} \delta_{d(x, x^\prime), \Delta} P(x|B)Q^{\pi}(x^\prime)P(x^\prime|B).
\end{equation}
Next, we may rewrite this equation as:
\begin{align}
   P(d(x,x^\prime)=\Delta|B) &= \left< \sum_{x} \delta_{d(x, x^\prime), \Delta} P(x|B)Q^{\pi}(x^\prime)\right>_{P(x^\prime|B) } \\
    &= \left< Q^{\pi}(x^\prime)\sum_{x} \delta_{d(x, x^\prime), \Delta} P(x|B)\right>_{P(x^\prime|B) }
\end{align}
We define $n_{\Delta}(x^\prime) = \sum_{x} \delta_{d(x, x^\prime), \Delta} P(x|B)$ to be the neighbourhood density around $x^\prime$ at distance $\Delta$. Assuming independence of neighbor density $n_{\Delta}(x^\prime)$ and selection factor $Q^{\pi}(x^\prime)$ we have
\begin{equation}
     P(d(x,x^\prime)=\Delta|B) = \left< Q^{\pi}(x^\prime)\right>_{P(x^\prime|B) } \left< n_{\Delta}(x^\prime)\right>_{P(x^\prime|B) }.
\end{equation}
By normalization the first term evaluates to $1$ so we are left with
\begin{equation}
     P(d(x,x^\prime)=\Delta|B) = \left< n_{\Delta}(x^\prime)\right>_{P(x^\prime|B) } =  \sum_{x, x^\prime} \delta_{d(x, x^\prime), \Delta} P(x|B) P(x^\prime|B) = p_C[P(X|B)](\Delta),
\end{equation}
therefore the likelihood ratio becomes
\begin{equation}
    \frac{P(d(x,x^\prime)=\Delta|\pi)}{P(d(x,x^\prime)=\Delta|B)} = \frac{p_C[P(X|\pi)](\Delta)}{p_C[P(X)](\Delta)}.
\end{equation}
Where once again we have used $ p_C[P(X)](\Delta) = p_C[P(X|B)](\Delta) $ to denote the background probability of coincidence.

 \newpage

\section{Feature interactions arising from data structure}

\subsection{Mixture of motifs model}
\label{app5}

In the following we explore in a minimal model how local feature relevancy and interaction information depend on the number of distinct binding solutions able to engage a particular epitope. We consider that for a given epitope $\pi$ there is a distribution of motifs/binding modes, $P(\Phi)$. For a given binding mode, $\Phi = \phi$, there is an associated set of specific sequences and sequence features, $P(X | \phi, \pi)$. The probability of drawing a particular feature $X = x$ from the distribution of sequences specific to $\pi$ is
\begin{equation}
    P(x | \pi) = \sum_{\phi} P(x| \phi, \pi) P(\phi),
\end{equation}
The coincidence probability for sequences drawn from $P(X | \pi)$ is
\begin{equation}
    p_C[P(X | \pi)] = \sum_{x}\left(\sum_{\phi} P(x | \phi, \pi) P(\phi)\right)^2
\end{equation}
In \cite{mayercallan} we show that such a mixture distribution may be expressed as the sum of a within-motif and a cross-motif term
\begin{equation}
    p_C[P(X | \pi)] = p_C[P(\phi)] \left< p_C[P(X| \pi, \phi)]\right>_{P(\phi | \phi_1=\phi_2=\phi)} + (1-p_C[P(\phi)]) \left< p_C[P(X| \pi, \phi_1), P(X| \pi, \phi_2)]\right>_{P(\phi_1, \phi_2 | \phi_1 \neq \phi_2)},
\end{equation}
where $\left< p_C[P(X| \pi, \phi)]\right>_{P(\phi | \phi_1=\phi_2=\phi)} = p_C[P(X| \pi, \Phi)]$ is the average probability of feature coincidence within each motif subset for motifs associated with epitope $\pi$ weighted by
\begin{equation}
P(\phi | \phi_1=\phi_2=\phi) = \frac{P(\phi)^2}{\sum_{\phi^\prime}P(\phi^\prime)^2 } = \rho_2(\phi).
\end{equation}
And $\left< p_C[P(X| \pi, \phi_1), P(X| \pi, \phi_2)]\right>_{P(\phi_1, \phi_2 | \phi_1 \neq \phi_2)}$ is the cross-motif coincidence probability with the probability of coincidence for two distributions defined in general as:
\begin{equation}
p_C[P_i(X), P_j(X)] = \sum_x P_i(x) P_j(x),
\end{equation}
with the average computed with respect to the following weighting factor
\begin{equation}
P(\phi_1, \phi_2 | \phi_1 \neq \phi_2) = \frac{P(\phi_1)P(\phi_2)}{1-\sum_{\phi^\prime}P(\phi^\prime)^2 }.
\end{equation}
We will assume that in general this cross-motif coincidence probability is far smaller than the within coincidence probability such that
\begin{equation}
    p_C[P(X | \pi)] \approx p_C[P(\phi)] p_C[P(X| \pi, \Phi)].
\end{equation}
To proceed, we will now assume that $p_C[P(\phi)] = 1/M$, where $M$ is number of motifs associated with a given epitope $\pi$. We will also assume that for a given $M$ there is a characteristic probability of coincidence $p_C[P(X | \Pi_M)]$ where $\Pi_M$ denotes the fact that we are considering epitopes with $M$ motifs. We use uppercase $\Pi$ to highlight that we are theorising that epitopes with equal number of motifs only vary in their coincidence probabilities for particular features due to finite sample size error and hence their true coincidence probabilities should be equal in the limit of infinite data. We will assume that $p_C[P(X| \pi, \Phi)] = p_C[P(X| \Pi_1)]$ such that it is the probability of feature coincidence for epitopes with a single motif. Therefore, for an epitope with $M$ motifs we obtain
\begin{equation}
    p_C[P(X | \Pi_M)] \approx \frac{1}{M}p_C[P(X| \Pi_1)] .
\end{equation}
The local relevancy of $X$ for an epitope with $M$ motifs is then
\begin{align}
    i_2(X, \Pi_M) &= \log \left( \frac{1}{M}\frac{p_C[P(X| \Pi_1)]}{p_C[P(X)]}\right) \nonumber \\
    &= \log \left( \frac{p_C[P(X| \Pi_1)]}{p_C[P(X)]}\right) - \log M \nonumber \\
    &= i_2(X, \Pi_1) - \log M ,
\end{align}
where $i_2(X, \Pi_1)$ is the characteristic local relevancy of feature $X$ for epitopes with a single motif. Therefore, the greater the number of motifs associated with a given epitope, the less locally relevant each feature is expected to become. With such a model we would expect the local relevancy of two features, $X$ and $Y$, for a given number of motifs to be related by the following equation
\begin{equation}
    i_2(X, \Pi_1) - i_2(X, \Pi_M) = i_2(Y, \Pi_1) - i_2(Y, \Pi_M).
\end{equation}
Such that
\begin{equation}
    i_2(X, \Pi_M) =  i_2(Y, \Pi_M) + i_2(X, \Pi_1) -i_2(Y, \Pi_1).
\end{equation}
Which is a relationship we observe in Figure \ref{results_5}. Now let us consider how we expect the local interaction information between two features to change with number of motifs
\begin{align}
    i_{2, int}(X, Y | \Pi_M) &= i_2([X, Y], \Pi_M) - i_2(X, \Pi_M)  - i_2(Y, \Pi_M) \\ 
    &= i_2([X, Y], \Pi_1) - i_2(X, \Pi_1)  - i_2(Y, \Pi_1) - \log M + 2\log M \\
    &= i_{2, int}(X, Y | \Pi_1) + \log M.
\end{align}
We expect the local interaction information between pairs of features to increase with the number of motifs associated to the particular eptiope. Therefore, we expect the interaction information between two features and their independent local relevancy for a given number of motifs to be related by the following expressions
\begin{align}
     i_{2, int}(X, Y | \Pi_M) - i_{2, int}(X, Y | \Pi_1) = i_2(X, \Pi_1) - i_2(X, \Pi_M) \\
     = i_2(Y, \Pi_1) - i_2(Y, \Pi_M).
\end{align}
So that
\begin{align}
     i_{2, int}(X, Y | \Pi_M) = - i_2(X, \Pi_M) + i_2(X, \Pi_1)  + i_{2, int}(X, Y | \Pi_1) \\
     = - i_2(Y, \Pi_M) + i_2(Y, \Pi_1) + i_{2, int}(X, Y | \Pi_1).
\end{align}
Which is a relationship observed in Figure \ref{results_5}. 

\subsection{False positives}
\label{app6}

We will briefly show that false positives, that is sequences which are not truly specific to a given epitope but yet appear in a specific TCR sample, lead to similar results to the mixture of motif models. Consider that a sample of sequences is taken in an experiment designed to obtain sequences specific to epitope $\pi$ and that a distribution of a particular feature $X$ is obtained, $P(X|\tilde{\pi})$. A fraction of features in this distribution are truly specific to $\pi$, $P(\pi)$, while a fraction of sequences are false positives, $P(\neg \pi)$. We will denote the subset of sequence features truly specific to epitope $\pi$ as $P(X|\pi)$ and the subset of sequence feature not specific to $\pi$ as $P(X|\neg \pi)$. The probability of coincidence for the distribution $P(X|\tilde{\pi})$ is
\begin{equation}
    p_C[P(X|\tilde{\pi})] = \sum_{x} \left( P(X|\pi) P(\pi) + P(X|\neg \pi) P(\neg \pi)\right)^2.
\end{equation}
Expanding out we get
\begin{align}
    p_C[P(X|\tilde{\pi})] &= \sum_{x} P(X|\pi)^2 P(\pi)^2 + \sum_{x} P(X|\neg \pi)^2 P(\neg \pi)^2 + 2\sum_{x} P(X|\pi)P(X|\neg \pi)P(\pi)P(\neg \pi) \\
    &= P(\pi)^2 p_C[P(X|\pi)] +  P(\neg\pi)^2 p_C[P(X|\neg \pi)] + 2P(\pi)P(\neg \pi) p_C[P(X|\pi), P(X|\neg \pi)].
\end{align}
We will assume that cross coincidences are rare, that coincidences are far more likely between the truly specific sequences and that the fraction of false positives is far smaller than the fraction of true positives such that the following expressions hold: $ P(\pi)^2 p_C[P(X|\pi)] \gg  2P(\pi)P(\neg \pi)p_C[P(X|\pi), P(X|\neg \pi)]$ and $P(\pi) p_C[P(X|\pi)] \gg  P(\neg \pi) p_C[P(X|\neg \pi)]$. Therefore we obtain
\begin{equation*}
    p_C[P(X|\tilde{\pi})] \approx  P(\pi)^2 p_C[P(X|\pi)].
\end{equation*}
Now considering the local relevancy of $X$ computed from $p_C[P(X|\tilde{\pi})]$
\begin{align}
     i_2(X, \tilde{\pi}) &= \log \left(\frac{P(\pi)^2 p_C[P(X|\pi)]}{p_C[P(X)]} \right) \nonumber \\
     &= \left(\frac{p_C[P(X|\pi)]}{p_C[P(X)]} \right) + 2\log P(\pi) \nonumber \\
     &=  i_2(X, \pi) + 2\log P(\pi).
\end{align}
If $P(\pi) < 1$, then $\log P(\pi) < 0$ so the relevancy of $X$ will be lower when computed using distribution $P(X|\tilde{\pi})$ than when using $P(X|\pi)$. Similarly to \ref{app5}, we identify that when $P(\pi) < 1$ the local interaction information between two features $X$ and $Y$ will be increased
\begin{align}
    i_{2, int}(X, Y | \tilde{\pi})  &= i_2([X , Y], \tilde{\pi}) - i_2(X, \tilde{\pi} ) - i_2(Y, \tilde{\pi} ) \nonumber \\
     &= i_2([X , Y],  \pi) - i_2(X, \pi ) - i_2(Y, \pi ) + 2\log P(\pi) - 2\log P(\pi) - 2\log P(\pi) \nonumber \\
     &= i_{2, int}(X, Y | \pi) - 2\log P(\pi).
\end{align}
Therefore the relationships observed in Figure \ref{results_5} could also be produced by false positives. Analyses of how TCR clustering depends on background sequence space coverage could provide a potential avenue for distinguishing between both models in future work. \newpage

\section{Data analysis methods} \label{methods}

\subsection{Data collection and pre-processing}

TCR sequencing data from \citet{minervina2022sars} and \citet{dash2017quantifiable} was processed as follows: Data was deduplicated on the full nucleotide sequence level for each epitope in order to enable assessment of clonal convergence independently of clonal expansions within individual donors \cite{mayercallan}. Only TCRs with full CDR3, V gene and J gene annotations for both the $\alpha$ and $\beta$ chain were retained, and V and J gene names were standardized using tidytcells \cite{nagano2023tidytcells}. Furthermore, epitope specific sets were only retained in cases where at least one full TCR coincidence was observed. Background TCR sequence data was produced using the OLGA generation model \cite{sethna2019olga}. We generated $1000000$ $\alpha$ and $\beta$ chain sequences and then randomly paired these to produce paired chain background data. The final number of sequences retained is detailed in Table \ref{raw_data}.

\begin{table}[h!]
\begin{center}
\begin{tabular}
{ |p{3cm}||p{3cm}|p{3cm}|p{3cm}|p{3cm}|  }
 \hline
 Epitope ID & Data set & Epitope & HLA & Sequence counts\\
 \hline
    & OLGA  &N/A& N/A &1000000\\
 
 1 & Minervina & NQKLIANQF & HLA-B*15:01 & 148 \\
 2 & Minervina & DTDFVNEFY & HLA-A*01:01 & 88 \\
 3 & Minervina & LLYDANYFL & HLA-A*02:01 &53 \\
 4 & Minervina & PTDNYITTY & HLA-A*01:01 & 155 \\
 5 & Minervina & YLQPRTFLL & HLA-A*02:01 & 288 \\
 6 & Minervina & FTSDYYQLY & HLA-A*01:01 &450 \\
 7 & Minervina & ALSKGVHFV & HLA-A*02:01 & 197 \\
 8 & Minervina & LTDEMIAQY & HLA-A*01:01 &398 \\
 9 & Minervina & TTDPSFLGRY & HLA-A*01:01 &1909 \\
 10 & Dash &   NLVPMVATV  & HLA-A*02:01 &67  \\
 11 & Dash &   GLCTLVAML  & HLA-A*02:01 &92  \\
  12 & Dash &   GILGFVFTL   & HLA-A*02:01 &249  \\
 \hline
\end{tabular}
\caption{TCR sequence data used in this study. The first column shows the epitope ID used throughout figures.}
\label{raw_data}
\end{center}
\end{table}

\subsection{Statistical analysis}

Coincidence probabilities and their variance were computed using unbiased estimators described in \cite{tiffeau2023unbiased}. Conditional probabilities of coincidence were averaged over epitopes using Eq. \ref{pccond} with an equal weighting assigned to each epitope specific subset. Background entropies of the paired chain TCRs were computed as the sum of $\alpha$ and $\beta$ chain entropies, exploiting the independent pairing of chains in the computationally constructed background. Mutual information scores and higher order statistics such as synergy were then computed from these entropies using Eqs. \ref{mieq}, \ref{local_mieq} and \ref{syn}. Orthogonal distance regression was performed as implemented in Scipy \cite{2020SciPy-NMeth}.

\begin{table}[h!]
\begin{center}
\begin{tabular}
{ |p{3cm}|p{3cm}|p{8cm}|}
 \hline
 Name & Version & Link\\
 \hline
 Tidytcells \cite{nagano2023tidytcells} & 2.00 & \url{https://pypi.org/project/tidytcells/}\\
 Pyrepseq \cite{mayercallan} & 1.2.1 & \url{https://pypi.org/project/pyrepseq/}\\
 OLGA  \cite{sethna2019olga} & 1.2.4 & \url{https://github.com/statbiophys/OLGA} \\
 Scipy  \cite{2020SciPy-NMeth} & 1.11.3 & \url{https://pypi.org/project/scipy/} \\
 \hline
\end{tabular}
\caption{Software packages used in this study.}
\label{software}
\end{center}
\end{table}

\newpage
\section*{Supplementary Figures}

\begin{figure*}[h!]
\includegraphics[width=\textwidth]{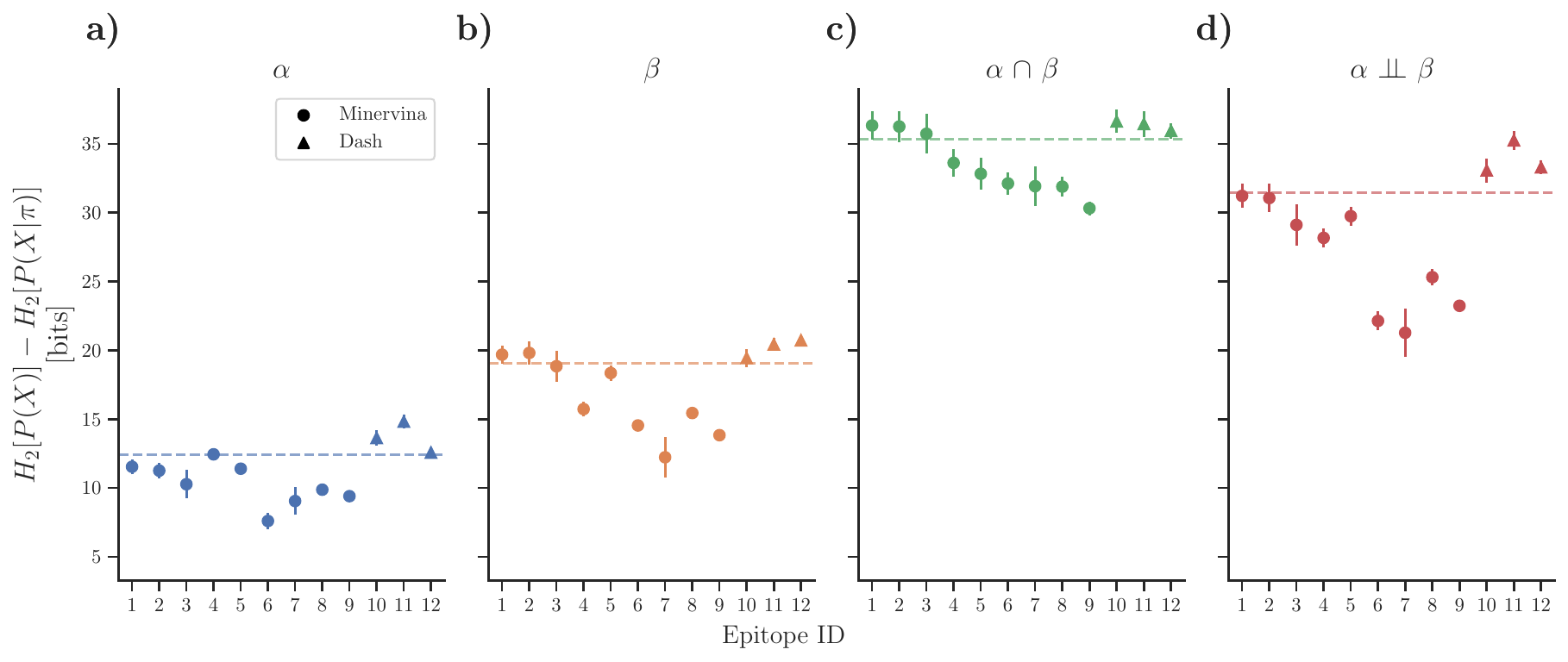}
\caption{\textbf{Coincidence entropy reduction on the per-epitope level.} 
Coincidence entropy reduction between specific and background TCRs for the \textbf{a)} $\alpha$ and \textbf{b)} $\beta$ chains. \textbf{c)} Local relevancy of the full paired TCR chain. $X \cap Y$ represents joint consideration of features $X$ and $Y$ here and in the following. \textbf{d)} Sum of local relevancies of the $\alpha$ and $\beta$ chain. $X \perp \!\!\! \perp Y$ represents separate consideration of features $X$ and $Y$ here and in the following. Epitopes are ordered within each dataset by their full entropy change. Dashed lines indicate the average entropy change shown in Figure \ref{results_1}. Although a large amount of variability between various epitopes not captured by their associated uncertainties is observed, every epitope shows a decrease in both $\alpha$ and $\beta$ chain entropy. For every epitope, the $\alpha$ and $\beta$ chain are therefore informative features. The change in entropy of the full TCR sequence $\alpha\beta$ is in all cases greater than the sum of the entropy changes of the $\alpha$ and $\beta$ chains taken independently, suggesting that synergy between $\alpha$ and $\beta$ chain information is a general phenomenon.}
\label{results_2}
\end{figure*}

\begin{figure*}[h!]
\includegraphics[width=\textwidth]{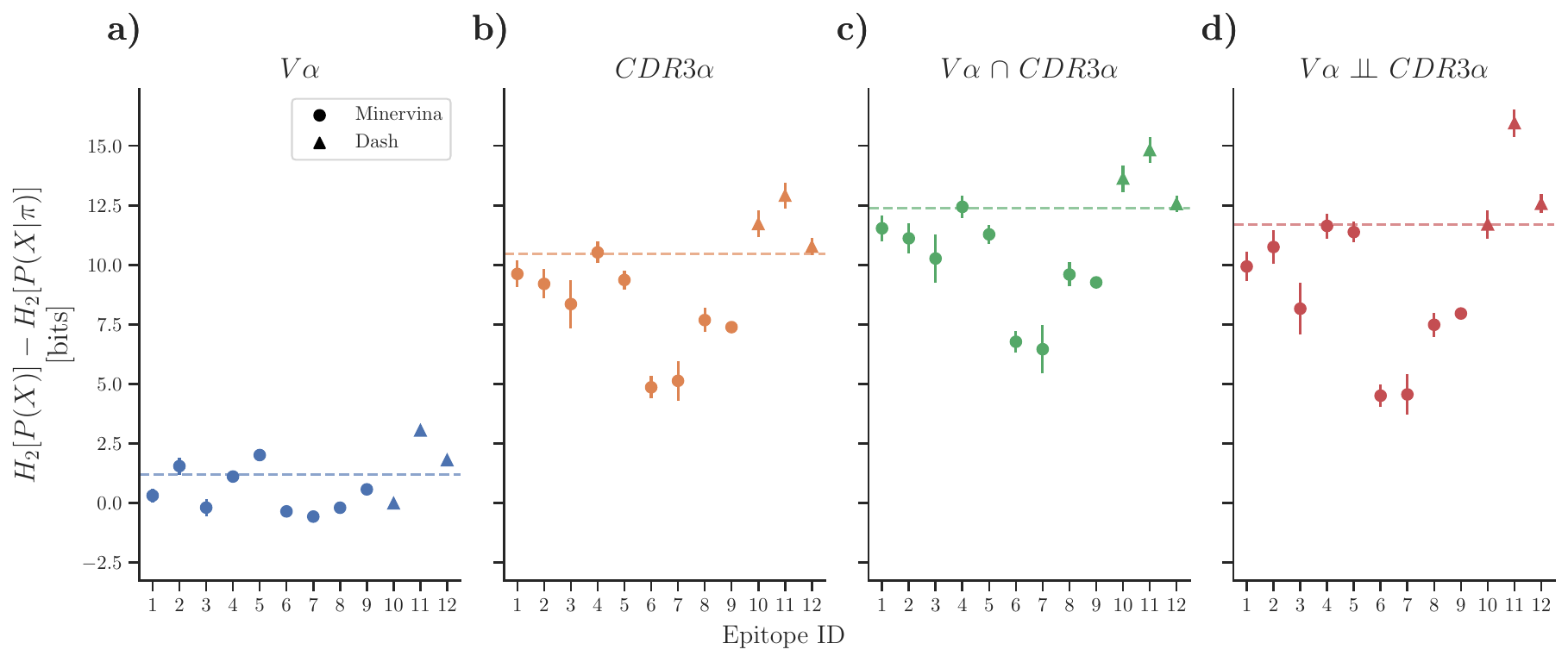}
\caption{\textbf{Coincidence entropy reduction for CDR3$\alpha$ and V$\alpha$.} Both the V$\alpha$ and CDR3$\alpha$ contributions to specificity vary across epitopes, with some epitopes showing no detectable restriction of V$\alpha$ gene choice. Further details as described for Fig.~\ref{results_2}.}
\label{results_va}
\end{figure*}

\begin{figure*}[h!]
\includegraphics[width=\textwidth]{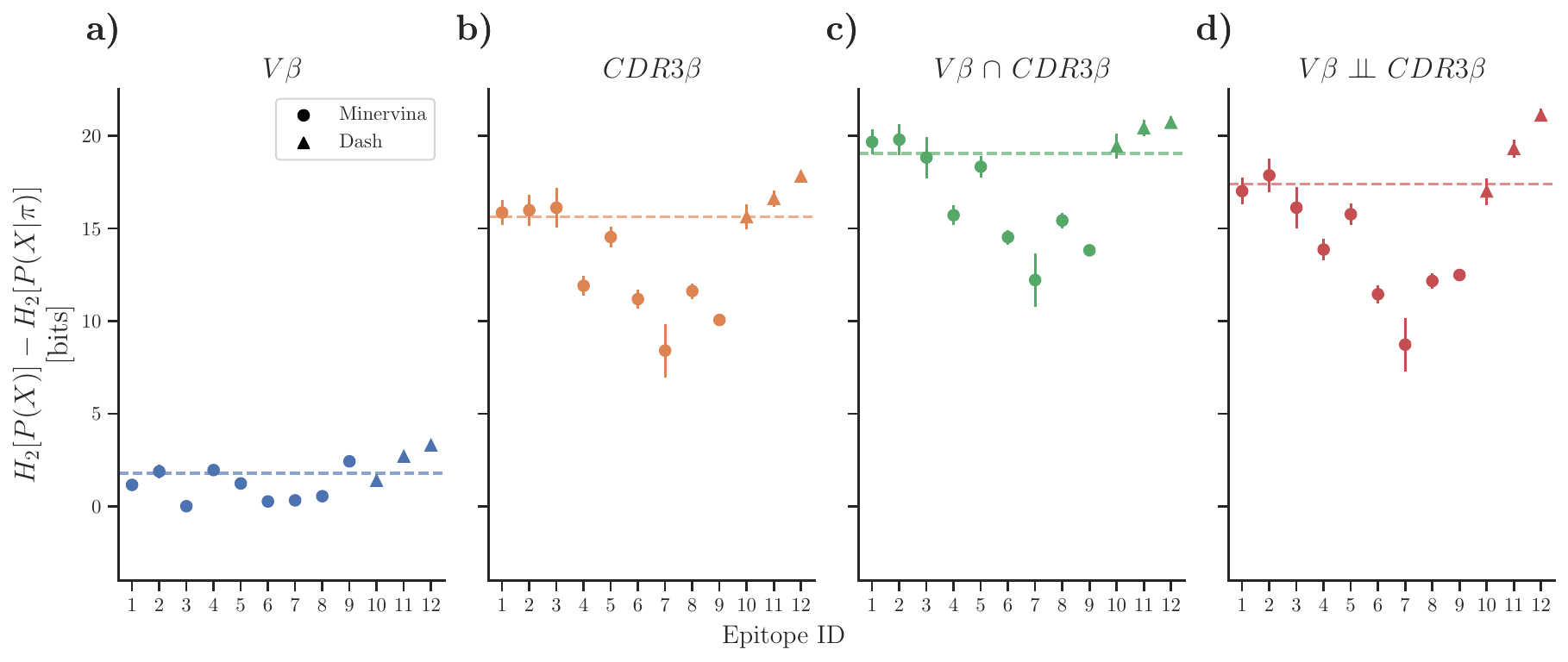}
\caption{\textbf{Coincidence entropy reduction for CDR3$\beta$ and V$\beta$.} Both the V$\beta$ and CDR3$\beta$ contributions to specificity vary across epitopes. Most epitopes show small, but positive restriction of V$\beta$ gene choice. Further details as described for Fig.~\ref{results_2}.}
\label{results_vb}
\end{figure*}

\begin{figure*}[h!]
\includegraphics[width=\textwidth]{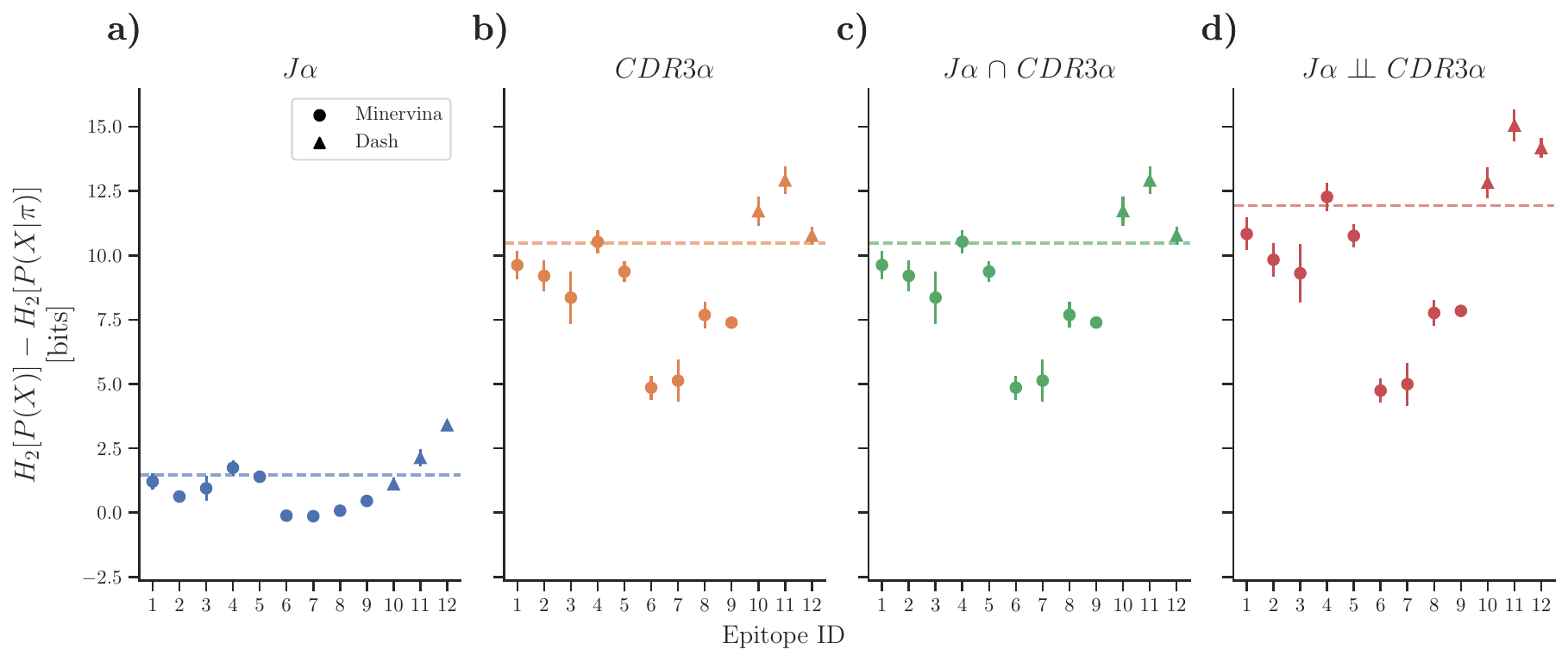}
\caption{\textbf{Coincidence entropy reduction for CDR3$\alpha$ and J$\alpha$.} Comparison of panel c and d reveals negative interaction information between J$\alpha$ and CDR3$\alpha$ regions suggesting that the CDR3$\alpha$ makes the J$\alpha$ redundant. Further details as described for Fig.~\ref{results_2}.}
\label{results_ja}
\end{figure*}

\begin{figure*}[h!]
\includegraphics[width=\textwidth]{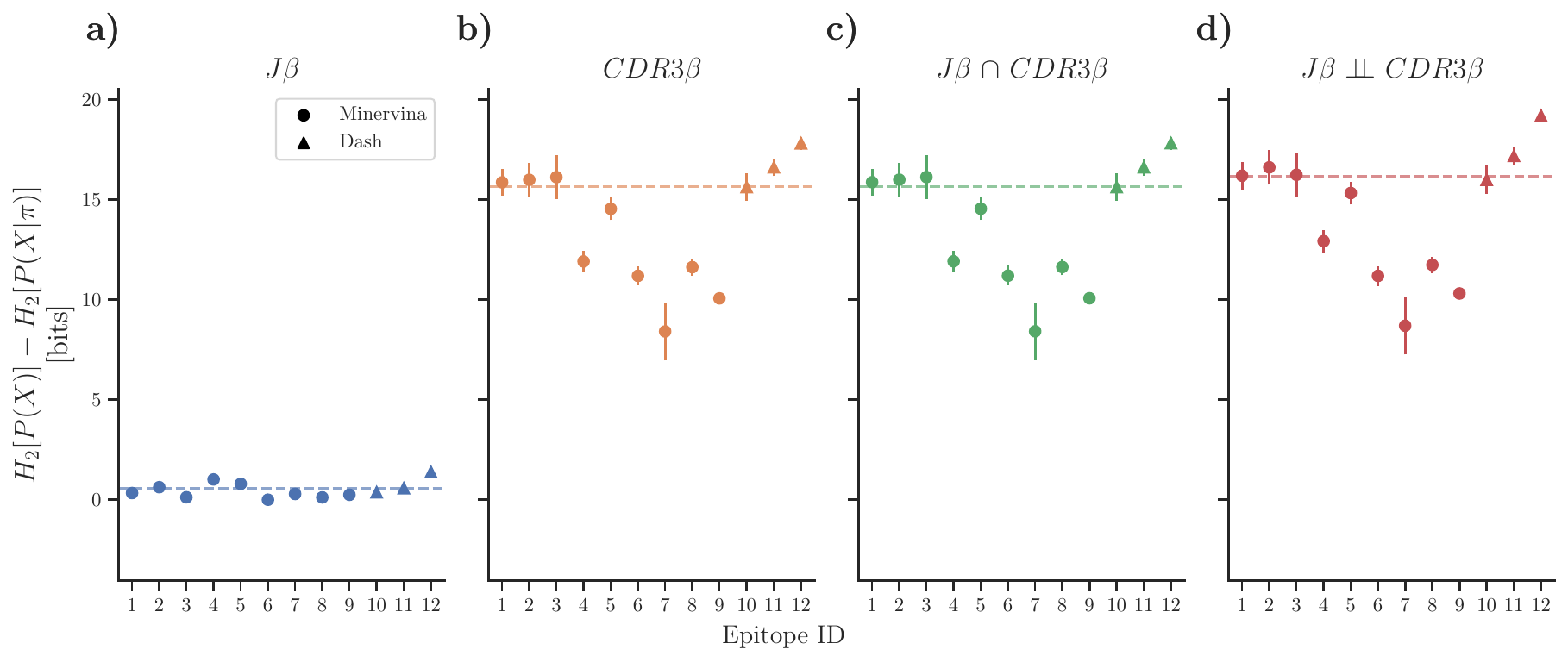}
\caption{\textbf{Coincidence entropy reduction for CDR3$\beta$ and J$\beta$.} Comparison of panel c and d reveals negative interaction information between J$\beta$ and CDR3$\beta$ regions suggesting that the CDR3$\beta$ makes the J$\beta$ redundant. Further details as described for Fig.~\ref{results_2}.}
\label{results_jb}
\end{figure*}

\begin{figure*}[h!]
\includegraphics[width=\textwidth]{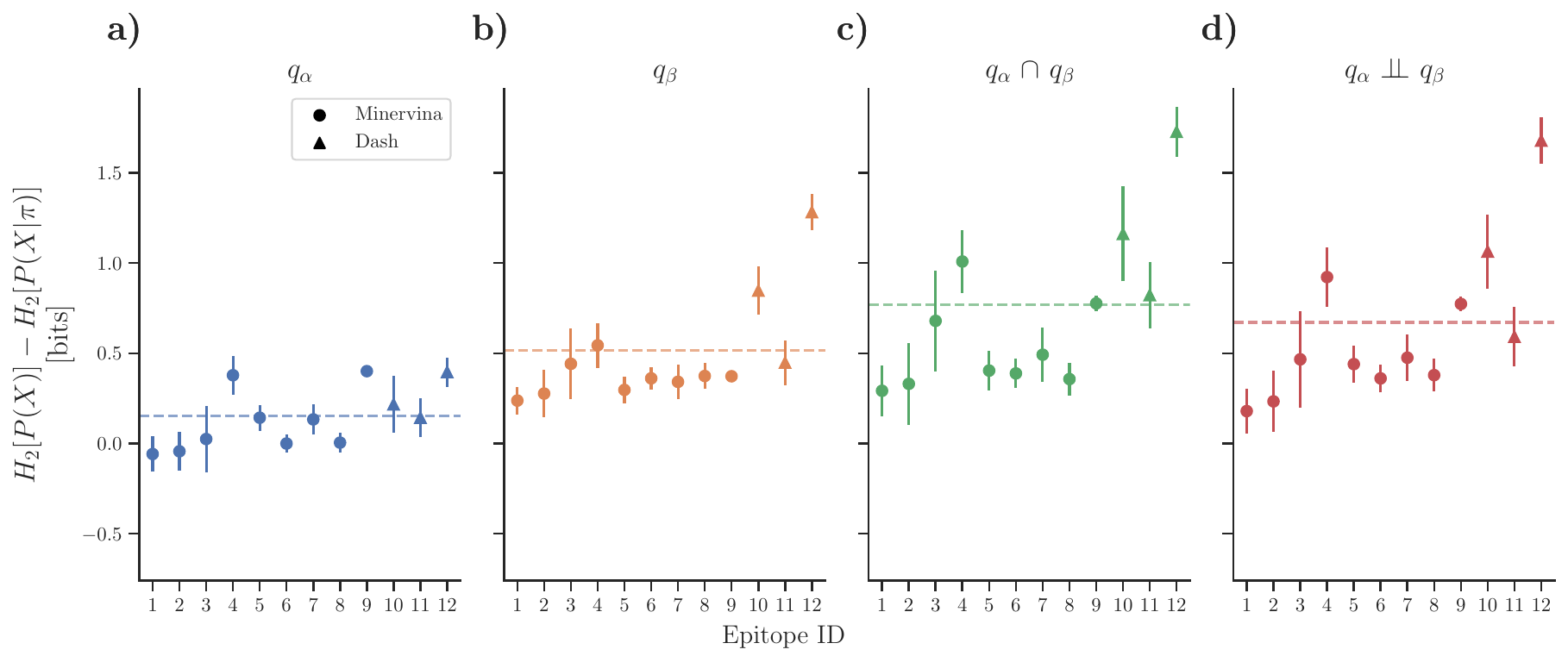}
\caption{\textbf{Coincidence entropy reduction for CDR3 net charge across epitopes.} While $\alpha$ chain charge has variable relevancy and in some cases is not informative, $\beta$ chain charge is consistently informative across epitopes. Further details as described for Fig.~\ref{results_2}.}
\label{results_3} 
\end{figure*}

\begin{figure*}[h!]
\includegraphics[width=\textwidth]{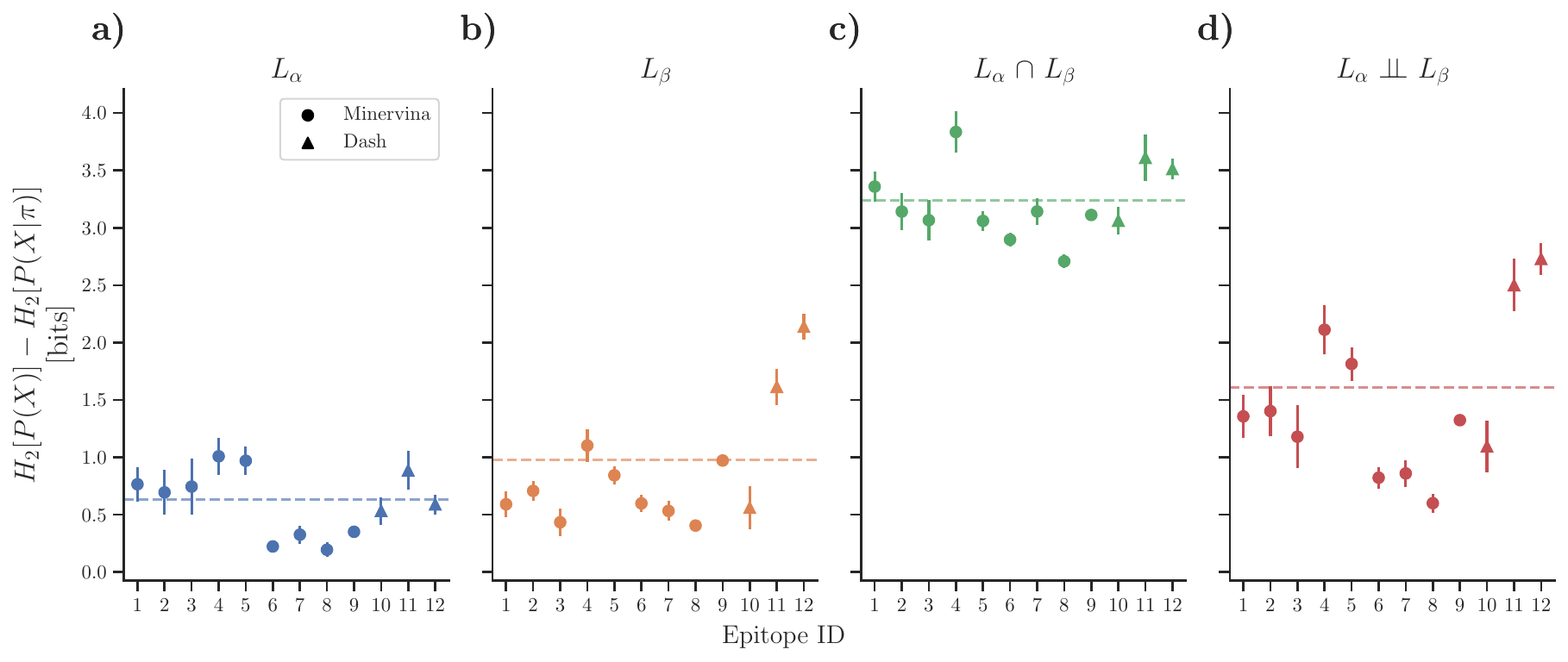}
\caption{\textbf{Coincidence entropy reduction for CDR3 length across epitopes.} The CDR3 length of both the $\alpha$ and $\beta$ chains are relevant features for all epitopes. There is substantial $\alpha$ and $\beta$ length pairing as all epitopes display positive interaction information (synergy). Further details as described for Fig.~\ref{results_2}.}
\label{results_4}
\end{figure*}

\begin{figure*}[h!]
\includegraphics[width=\textwidth]{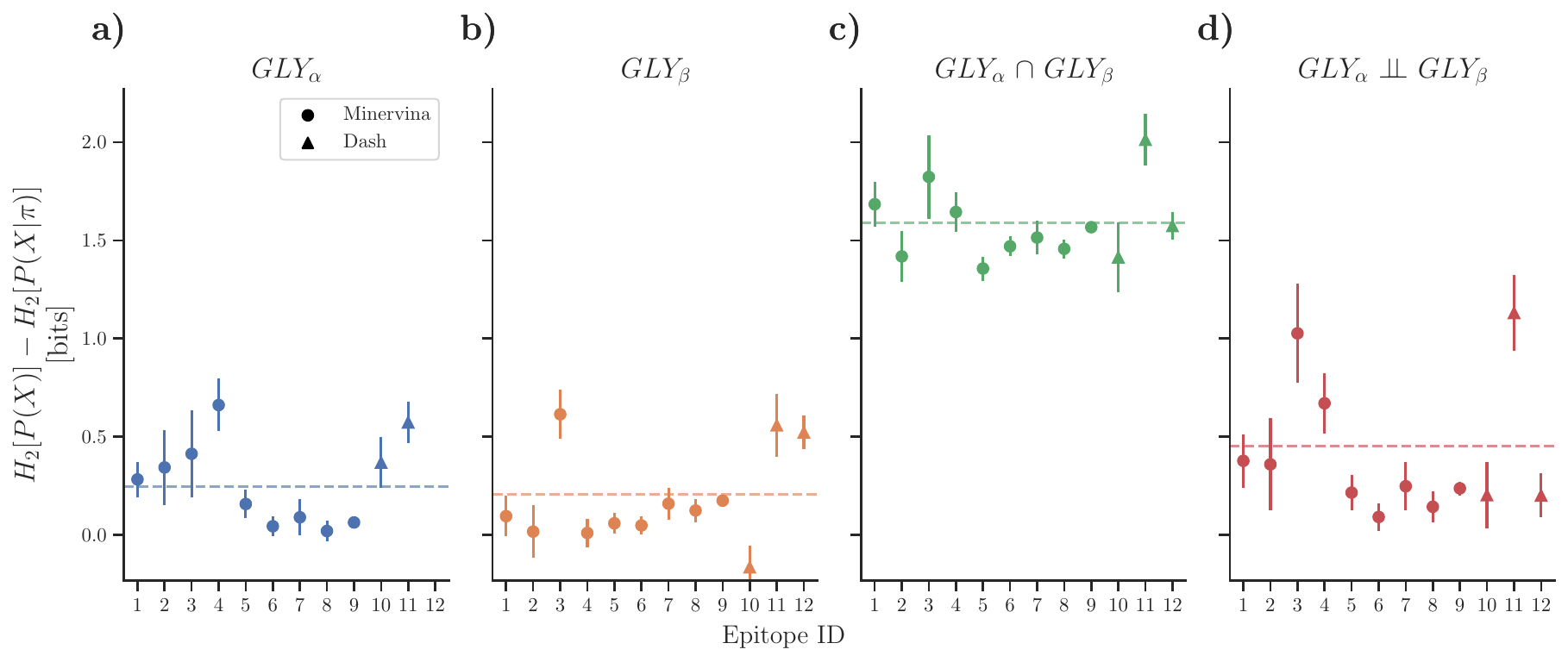}
\caption{\textbf{Coincidence entropy reduction for CDR3 glycine content across epitopes.} CDR3 glycine content, defined as the number of glycine residues within the CDR3, is only weakly relevant when considered on the single chain level, but there is substantial synergy. Further details as described for Fig.~\ref{results_2}.}
\label{results_gly}
\end{figure*}

\end{document}